\documentclass[12pt,epsf,amssymb]{article} \textheight =23 truecm
\def\lsim{\raise0.3ex\hbox{$<$\kern-0.75em\raise-1.1ex\hbox{$\sim$}}}
\def\gsim{\raise0.3ex\hbox{$>$\kern-0.75em\raise-1.1ex\hbox{$\sim$}}}
\def\noi{\noindent}  \def\bea{\begin{eqnarray}}
\def\eea{\end{eqnarray}} \def\beq{\begin{equation}}
\def\eeq{\end{equation}} 
\def\beeq{\begin{eqnarray}} \def\eeeq{\end{eqnarray}} \def\R{ {\rm R
\kern -.31cm I \kern .15cm}} \def\C{ {\rm C \kern -.15cm \vrule
width.5pt \kern .12cm}} \def\Z{ {\rm Z \kern -.27cm \angle \kern
.02cm}} \def\N{ {\rm N \kern -.26cm \vrule width.4pt \kern .10cm}}
\def\1{{\rm 1\mskip-4.5mu l} }

\usepackage{graphics} \usepackage{graphicx} \usepackage{epsfig}

\begin{document}

\begin{center} 

\vskip 3 truemm

{\Large \bf Remarks on the penguin decay $B_s \to \phi \phi$\par
with prospects for FCCee} \par

\par \vskip 5 truemm

 R. Aleksan $^a$ and L. Oliver $^b$

\vskip 2 truemm

$^a$ {\it IRFU, CEA, Universit\' e Paris-Saclay,
91191 Gif-sur-Yvette Cedex, France}

\vskip 2 truemm

$^b$ {\it IJCLab, P\^ole Th\'eorie, CNRS/IN2P3 et Universit\'e Paris-Saclay\par 
b\^at. 210, 91405 Orsay, France}

\vskip 5 truemm

\end{center}

\begin{abstract}

We underline the theoretical interest of the vector-vector penguin decay $B_s \to \phi \phi$, very clean from the experimental point of view. The CP-violation asymmetry $A_{CP}^{mix}$ comes from the interference of mixing and decay $\lambda_{\phi \phi} = {q \over p} {\overline{A} \over A}$. In the Standard Model (SM) and in the Naive Factorization limit, the CP phase from the mixing ${q \over p}$ exactly cancels the CP phase from the decay ratio ${\overline{A} \over A}$. Therefore, this mode is suitable to look for possible New Physics (NP) because $A_{CP}^{mix}$ would directly indicate the departure from the SM in mixing. We estimate the deviation from this cancellation by analyzing possible small effects in the SM, using in particular the QCD Factorization scheme. We compare the theoretical expectation for $A_{CP}^{mix}$ to the measurement of LHCb, and the implications for NP. We pay also special attention to the transverse amplitude $h = -$, the longitudinal and transverse polarization fractions, and the interesting helicity-dependent observables $\lambda_{\phi \phi}^{h=0}$ and $\lambda_{\phi \phi}^{h=-}$. On the other hand, we make an estimation of the expected sensitivity at the future FCCee experiment for the CP phase and modulus of $\lambda_{\phi \phi}$. We find $\delta(\mid \lambda_{\phi \phi}\mid) = 0.004$ and $\delta(\phi_{\phi \phi}) = 0.009$ rad and, comparing to the LHCb data, we point out the expectations at FCCee in the search of NP.

\end{abstract}

\section{{\Large Introduction}}

The physics of $B$ meson decays into two vector mesons is very rich due to the polarization degrees of freedom. In this paper we consider the Penguin decay $B_s \to \phi \phi$ that has been already experimentally studied at LHCb \cite{LHCbPHIPHI}. This mode is very clean since the final state includes two narrow resonances producing altogether four charged kaons, which offers a clean experimental signature with very low background, provided the detector includes a good Particle Identification system.\par
On the other hand, this decay of a heavy meson into two light mesons is rather well understood in the SM within the QCD Factorization (QCDF) scheme.\par 

We will here concentrate on two main kinds of observables, namely the polarization fractions, and the CP asymmetries in the interference between mixing and decay, 
\beq
\label{1bise} 
\lambda_{\phi \phi} =  \left({q \over p}\right)_{B_s} {A(\overline{B}_s \to \phi \phi) \over A(B_s \to \phi \phi)}
\eeq

\noi that occur in the time-dependent decay, in particular $A_{CP}^{mix} \simeq {\rm Im} \lambda_{\phi \phi}$.\par
In Section 2 we describe the puzzle of the polarization fractions in Penguin decays like $\overline{B}_s \to \phi \phi$, as compared to tree decays.
In Section 3 we describe the time-dependent decay of $B_s \to \phi \phi$, underlying the interesting observables and pointing out the result $A_{CP}^{mix} = 0$ with top dominance in mixing and Naive Factorization in decay. In Section 4, we estimate the deviations from this naive result by analyzing possible small effects, e.g. the contribution of other quarks in the box diagram and corrections of $O(\Gamma_{12} / M_{12})$ to mixing, and the ratio of decay amplitudes $\overline{A}/A$ beyond Naive Factorization (NF) within the QCDF scheme applied to $B$ meson decays into two light vector mesons \cite{KAGAN,BENEKE-2007,CHENG-YANG,BUCHALLA}. We gather the calculations for $B_s \to \phi \phi$ and give an estimation of  $A_{CP}^{mix}$, to see by how much it differs from its vanishing in the NF calculation. In Section 5 we compare to the   LHCb data and its implications for the search of New Physics (NP) in $B_s-\overline{B}_s$ mixing. In Section 6 we estimate the expected sensitivity of FCCee for the modulus and phase of $\lambda_{\phi \phi}$ and the implications for NP. In Section 7 we further comment in detail on the polarization-dependent rates and CP violation in the QCDF scheme. Finally, in Section 8 we make some caution remarks, and we conclude. In the Appendices we give useful formulas on the Wilson expansion (Appendix A), on the transverse $h = -$ amplitude (Appendix B) and on the annihilation for the longitudinal $h = 0$ amplitude within QCDF (Appendix C).

\section{Polarization fractions for the decay $\overline{B}_s \to \phi \phi$}
To be definite, let us write down the helicity amplitudes
\beq
\label{1e} 
\overline{A}_L = A[\overline{B} \to V_1(0) V_2(0)]\ , \qquad \qquad \overline{A}_\pm = A[\overline{B} \to V_1(\pm) V_2(\pm)]
\eeq

\noi where $V_1$ and $V_2$ are respectively the emitted and the produced vector mesons.\par 

Then the transversity amplitudes read
\beq
\label{2e} 
\overline{A}_\parallel = {1 \over \sqrt{2}} (\overline{A}_+ + \overline{A}_-) \ , \qquad \qquad \overline{A}_\perp = {1 \over \sqrt{2}} (\overline{A}_+ - \overline{A}_-)
\eeq

\noi with the corresponding transversity rate fractions $f_L , f_\parallel$ and $f_\perp$, satisfying $f_L + f_\parallel + f_\perp = 1$.\par

Since the final quark is dominantly left-handed because of the $V-A$ structure of the Standard Model (SM), heavy quark symmetry implies the hierarchy
\beq
\label{2-2e} 
\overline{A}_L : \overline{A}_- : \overline{A}_+ = 1 : {\Lambda_{QCD} \over m_b} : \left({\Lambda_{QCD} 
\over m_b}\right)^2
\eeq

\vskip 3 truemm

As underlined in detail by Beneke et al. \cite{BENEKE-2007}, the transverse amplitude $\overline{A}_-$ is suppressed by a factor $m_{V_2}/m_B$ relative to $\overline{A}_L$, and the axial and vector contributions to $\overline{A}_+$ cancel in the heavy quark limit, implying the hierarchy (\ref{2-2e}).\par

The limit
\beq
\label{3e} 
\overline{A}_+ = 0
\eeq

\noi implies \cite{BENEKE-2007},
\beq
\label{4e} 
f_\parallel = f_\perp
\eeq

The hierarchy (\ref{2-2e}) points out to 
\beq
\label{4-1bise} 
f_\parallel \simeq f_\perp << f_L
\eeq

\vskip 3truemm

\includegraphics[scale=0.6]{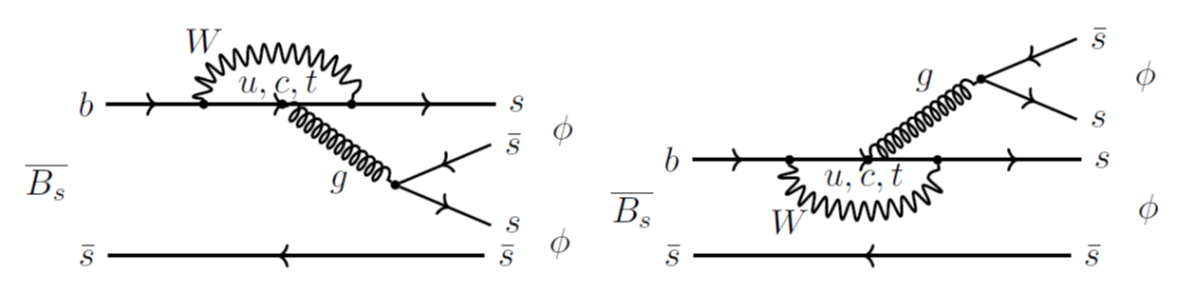}

Fig. 1. A few perturbative diagrams contributing to the decay $B_s \to \phi \phi$, where $g$ denotes one or several gluons with the right quantum numbers. 

\vskip 3truemm
The decay mode $B_s \to \phi \phi$ is a penguin mode, as shown in Fig. 1 in terms of perturbative diagrams, and in Fig. 2 in terms of local Wilson Penguin operators. 

\vskip 10truemm

\qquad \qquad \qquad \qquad  \includegraphics[scale=0.6]{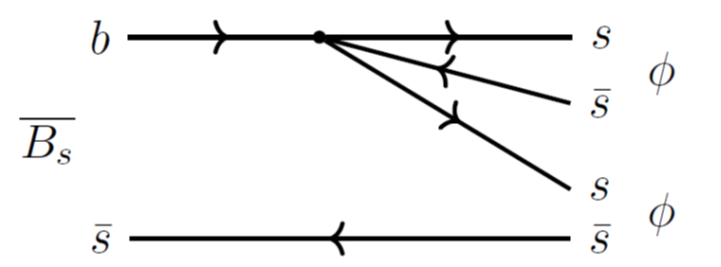}

\vskip 3truemm

Fig. 2. One diagram in terms of local Wilson Penguin operators $O_i^{Penguin}$ contributing to the decay $B_s \to \phi \phi$ .

\vskip 3truemm

It is very important to underline that the data for penguin-dominated $B$ decays are in conflict with the expected hierarchy (\ref{4-1bise}) \cite{BELLE05, BABAR06, BABAR07} if one uses Naive Factorization, that predicts a large longitudinal fraction $f_L \sim 1$, in consistency with the data for tree decay modes, e.g. $\overline{B}_d \to \rho \rho$. We illustrate this point with the polarization data for $B_s \to \phi \phi$ \cite{LHCbPHIPHI} in Table 1. 

\vskip 7truemm

{\small
\begin{center}
\begin{tabular}{|c|c|c|c|c|}
\hline 
Decay mode & $\rm{BR}\ (\times 10^{-6})$ & $f_L$ & $f_\parallel$ & $f_\perp$\\ \hline \hline 
$\overline{B}_s \to \phi \phi$ & $18.5 \pm 1.7$ &
$0.38 \pm 0.01$ & $ $ & $ 0.29 \pm 0.01$ \\ \hline
\end{tabular}
\end{center}}

\begin{center}
\noi Table 1. Data on the rate and polarization fractions for the decay mode $\overline{B}_s \to \phi \phi$.
\end{center}

Among the Penguin transitions, $\overline{B}_s \to \phi \phi$ appears as a privileged mode, since one has for it a wealth of data, $f_L, f_\perp$ and from Table 1 and the normalization condition, one gets $f_\parallel = 0.33 \pm 0.02$, that is rather precisely determined and consistent within errors with $f_\perp$, in agreement with the expectation following from (\ref{2-2e},\ref{4-1bise}). Also, $\overline{B}_s \to \phi \phi$ being a decay of the heavy meson into two light mesons, the theoretical QCDF framework applies.\par

The QCDF scheme includes NLO non-factorizable corrections of $O(\alpha_s)$, and the corresponding {\it effective} Wilson coefficients $a_i^h$ are {\it helicity-dependent}. One could then expect that QCDF could eventually explain the gap between the data of Table 1 and the naive hierarchy (\ref{4-1bise}). This is indeed the trend, as has been shown by the pionnering papers of Kagan \cite{KAGAN}, Beneke et al. \cite{BENEKE-2007} and Cheng and Yang \cite{CHENG-YANG}.\par 
For the particular case $\overline{B}_s \to \phi \phi$, we expose the main lines of the QCDF calculation of the polarization fractions in Section 7 and Appendices B, C where, following the papers by Beneke et al. \cite{BENEKE-2007} and by Bartsch et al. \cite{BUCHALLA}, we confirm the quantitative agreement with the experimental polarization fractions and rate.
We also use below QCDF for the calculation of the time-dependent CP violation \cite{BUCHALLA}.\par 

\section{Time-dependent CP asymmetries in $\overline{B}_s \to \phi \phi$}

For a decay of $\overline{B}_s$ into a final state $f$, the time-dependent rate reads, up to small terms of $O(\Gamma_{12}/M_{12})$\ \cite{NIERSTE-2000}, 
$$\Gamma(\overline{B}_s(t) \to f) =\ \mid A_f\mid^2 {1 + \mid \lambda_f\mid^2 \over 2}\ e^{-\Gamma t}$$
\beq
\label{5-3tere} 
\times \left[\cosh {\Delta \Gamma t \over 2} - A^{dir}_{CP} \cos (\Delta m t) + A_{\Delta \Gamma} \sinh {\Delta \Gamma t \over 2} - A_{CP}^{mix} \sin (\Delta m t) \right]
\eeq

\noi and for the case $\Gamma(B_s(t) \to f)$ the sign changes in front of the $A^{dir}_{CP}, A^{mix}_{CP}$ terms.\par 
In (\ref{5-3tere}), $\Delta m = m_H -m_L$ is the mass difference between the $B_s$ eigenstates ($H$, $L$ stand for heavy and light states), $\Gamma = {\Gamma_H+\Gamma_L \over 2}$, $\Delta \Gamma = \Gamma_L-\Gamma_H$ and $A_f$ is the decay amplitude for $B_s \to f$.
One has the coefficients
\beq
\label{5-3e} 
 A^{dir}_{CP} = {1 - \mid \lambda_f\mid^2 \over 1 + \mid \lambda_f\mid^2}\ , \qquad  A_{CP}^{mix} = - {2 {\rm \ Im} {\lambda_f} \over 1 + \mid \lambda_f\mid^2} \ , \qquad A_{\Delta \Gamma} = - {2 {\rm \ Re} {\lambda_f} \over 1 + \mid \lambda_f\mid^2}
\eeq

\noi that satisfy $\mid A^{dir}_{CP} \mid^2 + \mid A_{CP}^{mix} \mid^2 + \mid A_{\Delta \Gamma} \mid^2\ = 1$.

In eqns. (\ref{5-3tere},\ref{5-3e}) the crucial quantity $\lambda_f$ is
\beq
\label{5-1bise} 
\lambda_f = \left( {q \over p}\right)_{B_s} {A(\overline{B}_s \to f) \over A(B_s \to f)}
\eeq

Neglecting terms of $O(\Gamma_{12}/M_{12})$ and assuming top dominance in the box diagram for the mixing, $(q/p)_{B_s}$ is a pure phase. Moreover, in the case of a CP eigenstate, if in the amplitude one has a dominant product of CKM matrix elements, ${A(\overline{B}_s \to f_{CP}) \over A(B_s \to f_{CP})}$ is also a pure phase. Under these conditions, $\lambda_{f_{CP}}$ is a pure phase.\par

Note however that if two different weak factors, affected by two different strong phases, contribute to the decay amplitude, the ratio ${A(\overline{B}_s \to {f_{CP}}) \over A(B_s \to {f_{CP}})}$ and a fortiori $\lambda_{f_{CP}}$, are not pure phases. This is the case within the QCDF framework since in the amplitude $A(\overline{B}_s \to {f_{CP}})$ the $u$ and $c$ quarks in the Penguin acquire different NLO QCD corrections.\par

In the case of two vector mesons like $\phi \phi$ the interference between mixing and decay now depends on the polarization,  
\beq
\label{5-1e} 
\lambda_{\phi \phi}^{(k)} = \left({q \over p}\right)_{B_s} {A(\overline{B}_s \to \phi \phi, k) \over A(B_s \to \phi \phi, k)} = \eta_k \mid\lambda_{\phi \phi}^{(k)} \mid e^{-i\phi_{\phi \phi}^{(k)}}
\eeq
\noi where $k$ labels the polarization, longitudinal $k = L$, transverse parallel $k =\ \parallel$ and transverse perpendicular $k =\ \perp$, and $\eta_k$ is the corresponding CP eigenvalue, $\eta_L= \eta_\parallel = +1$ and $\eta_\perp = -1$.\par

\subsection{LHCb polarization-independent fit}

LHCb finds, in a polarization-independent fit \cite{LHCbPHIPHI},
\beq
\label{5-2e} 
\lambda_{\phi \phi}^{(k)} =\ \eta_k \mid \lambda_{\phi \phi} \mid e^{-i \phi_s^{s\overline{s}s}}
\eeq
\beq
\label{5-2-1e} 
\phi_s^{s\overline{s}s} = -0.073 \pm 0.115 \ , \qquad \qquad \mid\lambda_{\phi \phi}\mid\ = 0.99 \pm 0.05
\eeq

Although the quantity (\ref{5-2e}) is polarization-dependent because of the signs $\eta_k$, in the so-called "polarization-independent fit" of LHCb, the modulus $\mid \lambda_{\phi \phi} \mid$ and the phase $\phi_s^{s\overline{s}s}$ are assumed to be independent of the polarization. In Appendix B we compute in the QCDF scheme the quantities defined by (\ref{5-1e}), the moduli $\mid\lambda_{\phi \phi}^{(k)} \mid$ and the phases $\phi_{\phi \phi}^{(k)}$ for $k = L, \parallel, \perp$ and check that the hypothesis of a "polarization-independent fit" makes sense in a qualitative way.\par

To compare the theory with the LHCb data (\ref{5-2-1e}) we write now (\ref{5-1e}) with the notation
\beq
\label{5-1-1e} 
\lambda_{\phi \phi}^{(k)} = \eta_k \lambda_{\phi \phi}
\eeq
\beq
\label{5-1-2e} 
\lambda_{\phi \phi} =\ \mid \lambda_{\phi \phi} \mid e^{-i\phi_{\phi \phi}}
\eeq

Neglecting $\Delta \Gamma$ one finds the CP asymmetry,

$$A_{CP}^{mix}(t) = {\Gamma(\overline{B}_s(t) \to \phi \phi) - \Gamma(B_s(t) \to \phi \phi) \over \Gamma(\overline{B}_s(t) \to \phi \phi) + \Gamma(B_s(t) \to \phi \phi)}$$
\beq
\label{5-5e} \simeq  (f_L + f_\parallel - f_\perp)\ {\rm Im}(\lambda_{\phi \phi}) \sin (\Delta M t) = (1 - 2f_\perp)\ {\rm Im}(\lambda_{\phi \phi}) \sin (\Delta M t)
\eeq

\vskip 3 truemm

 On the other hand, if we assume the limit (\ref{3e}), $\overline{A}_+ = 0$, the CP asymmetry $A_{CP}^{mix}(t)$ can be expressed in terms of just the longitudinal polarization fraction $1 - 2f_\perp \simeq f_L$.\par

In conclusion, concerning the mixing CP asymmetry $A_{CP}^{mix}$ in the case $\overline{B}_s \to \phi \phi$, assuming that the CP phase is the same for the different polarizations, as done in \cite{LHCbPHIPHI}, one needs only $f_\perp$, or to a rather good approximation $f_L$, both already mesured at LHCb, as shown in Table 1.

\subsection{$A_{CP}^{mix}$ for $\overline{B}_s \to \phi \phi$ with top dominance in mixing and Naive Factorization in decay}

With top dominance one gets for the mixing,
\beq
\label{9e} 
\left({q\over p}\right)_{B_s} \simeq - {M_{12}^* \over \mid M_{12} \mid } = - \sqrt{ M_{12}^* \over M_{12} } \simeq - \sqrt{{(V^*_{tb} V_{ts})^2 \over (V_{tb} V^*_{ts})^2}}= - {V^*_{tb} V_{ts} \over V_{tb} V^*_{ts}} 
\eeq

In the decay $B_s \to \phi \phi$ within NF the Penguin amplitude is proportional to $-V_{ts}^*V_{tb}$ with the coefficient $C_4 + C_3/N_c$, where the Wilson coefficients $C_i$ are given in Table 11. Then we have, with the convention $CP \mid B >\ = - \mid \overline{B} >$, 
\beq
\label{5-7e} 
\lambda_{\phi \phi} = \left({q \over p}\right)_{B_s} {A(\overline{B}_s \to \phi \phi) \over A(B_s \to \phi \phi)} \simeq \left( - {V_{tb}^* V_{ts} \over V_{tb} V_{ts}^*}\right) \left(- {V^*_{ts} V_{tb} \over V_{ts} V^*_{tb}} \right)= 1
\eeq

\noi and there is no CP violation in this limit.

\section{{\Large Corrections to the cancellation of the CP phase}}

\subsection{Corrections to mixing in the Standard Model}

We consider here possible small corrections to the cancellation of CP violation in the SM (\ref{5-7e}) and in Section 5.1 we will introduce possible physics beyond the SM in mixing.

\subsubsection{Contributions of other quarks to the box diagram}

\noi Unlike $K-\overline{K}$ mixing, that has large corrections from $ct, cc$ corrections to $tt$ in the box diagram loop, $B-\overline{B}$ mixing is dominated by the top. However, since in this latter case we are interested in the approximate cancellation between mixing and decay, let us compute the corrections to top dominance in mixing. Including $(t, c, u)$ and using unitarity $V_{ub}V_{ud}^* = - V_{cb}V_{cd}^* - V_{tb}V_{td}^*$, one obtains \cite{NIERSTE-2009},
\beq
\label{10e} 
\left({q\over p}\right)_{B_q} \simeq - \sqrt{{(V^*_{tb} V_{tq})^2 S(x_t) + 2 (V^*_{tb} V_{tq}) (V^*_{cb} V_{cq}) S(x_c,x_t) + (V^*_{cb} V_{cq})^2 S(x_c) \over (V_{tb} V^*_{tq})^2 S(x_t) + 2 (V_{tb} V^*_{tq}) (V_{cb} V^*_{cq}) S(x_c,x_t) + (V_{cb} V^*_{cq})^2 S(x_c)}}
\eeq

\noi From 
\beq
\label{11e} 
x_i = {m_i^2 \over M_W^2}\ , \qquad \qquad x_t \simeq 4.6\ , \qquad \qquad x_c \simeq 3.5 \times 10^{-4}
\eeq 

\noi one gets \cite{BUCHALLA-BURAS-1990}
$$S(x_t) = x_t \left[{1 \over 4} + {9 \over 4} {1 \over (1 - x_t)} - {3 \over 2} {1 \over (1-x_t)^2}\right] - {3 \over 2} \left( {x_t \over 1 - x_t} \right)^3 \log{x_t} \simeq 2.5 \ \ \ $$
\beq
\label{12e}
S(x_c) \simeq x_c \simeq 3.5 \times 10^{-4} \qquad \qquad \qquad \qquad \qquad \qquad \qquad \qquad \qquad \ \  
\eeq
$$S(x_c,x_t) = x_c \left[\log {x_t \over x_c} - {3 x_t \over 4(1-x_t)} - {3 x_t^2 \log x_t \over 4(1-x_t)^2}\right] \simeq 3 \times 10^{-3} \qquad \qquad$$

\vskip 5 truemm

\noi Finally one finds, taking into account both $(c, t)$ in the mixing, 
\beq
\label{13e}
\left({q\over p}\right)_{B_s}  = - 0.9994 - 0.0350 i
\eeq

\noi and, keeping only top dominance, one gets a very close result,
\beq
\label{14e}
\left({q\over p}\right)_{B_s}  = - 0.9994 - 0.0349 i
\eeq 

\noi Therefore, the correction to top dominance is only of $O(10^{-4})$, and we can very safely make the approximation
\beq
\label{10bise} 
\left({q\over p}\right)_{B_s} \simeq - {V^*_{tb} V_{ts} \over V_{tb} V^*_{ts}}
\eeq

\subsubsection{Corrections to mixing of $O(\Gamma_{12} / M_{12})$}

With the phase of $M_{12}$ in the limit of top dominance being 
\beq
\label{4.2-2e}
\phi_M = \arg M_{12} = \arg\left(V_{tb}V_{ts}^*\right)^2
\eeq

\noi the mixing parameter is given by \cite{NIERSTE-2000}, 
\beq
\label{4.2-3e}
\left({q\over p}\right)_{B_s} = - e^{- i \phi_M} \left( 1 - {a \over 2} \right)
\eeq

\noi where the parameter $a$ is given by
\beq
\label{4.2-4e}
a = {\mid \Gamma_{12} \mid \over \mid M_{12} \mid} \sin \phi 
\eeq

\noi and the phase $\phi$ is defined by
\beq
\label{4.2-5e}
{M_{12} \over \Gamma_{12}} = - {\mid M_{12} \mid \over \mid \Gamma_{12} \mid}\ e^{i \phi}
\eeq

\noi From 
\beq
\label{4.2-1e}
\Delta M = 2 \mid M_{12} \mid \ , \qquad \qquad \Delta \Gamma = 2 \mid \Gamma_{12} \mid
\eeq

\noi we have therefore the upper bound
\beq
\label{4.2-6e}
\mid a \mid\ \leq {\Delta \Gamma \over \Delta M} \simeq {0.091 \over 17.761} \simeq 5 \times 10^{-3}
\eeq

\noi From this bound we conclude that the correction to $\left({q\over p}\right)_{B_s}$ being the pure phase dominated by top exchange,
\beq
\label{4.2-7e}
\left({q\over p}\right)_{B_s} = - e^{- i \phi_M}
\eeq 

\noi is at most of $O(2.5 \times 10^{-3})$.

\subsection{QCD Factorization corrections to the ratio of decay amplitudes $A(\overline{B}_s \to \phi \phi) / A(B_s \to \phi \phi)$}

To go beyond the Naive Factorization calculation, the natural theoretical scheme is QCD Factorization (QCDF), that starts from short distance operators with their Wilson coefficients, including the Penguin and EW Penguin operators, given in Appendix A. Next, this scheme includes NLO $O(\alpha_s)$ Vertex corrections $V$, Penguin diagrams $P$, Hard Spectator diagrams $H$ and Weak Annihilation $A$ (Fig. 3).\par

\includegraphics[scale=0.55]{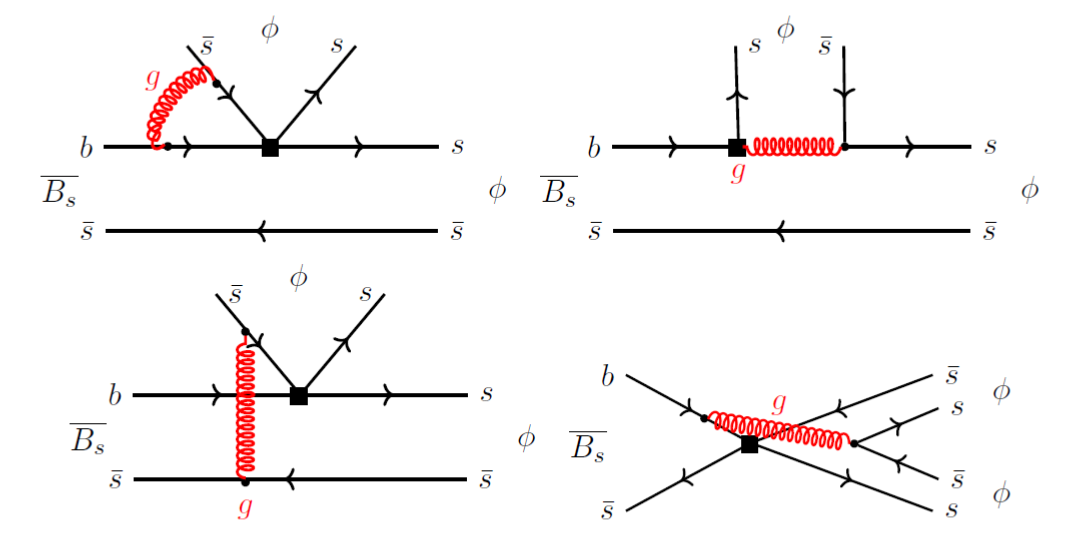}

Fig. 3. Typical NLO $\alpha_s$ corrections of the QCDF scheme. From left to right and from above to below : corrections of the types Vertex $V$, Penguin $P$, Hard Spectator $H$ and Annihilation $A$.

\vskip 3truemm

The Wilson coeffficients $C_i\ (i=1,... 10)$ at NLO \cite{BBNS-01}, for $\mu = m_b$, with $m_b(m_b) = 4.2\ \rm{GeV}$, $\Lambda_{\overline{MS}}^{(5)} = 0.225\ \rm{GeV}$, are given in Table 11 of Appendix A.\par

Schematically, the matrix elements have the structure \cite{BBNS-99,BBNS-00,BENEKE-NEUBERT},
$$<V_1 V_2\mid O_i \mid \overline{B} >\ = \left[F^{\overline{B} \to V_1} T_i^I \otimes f_{V_2} \Phi_{V_2} + (V_1 \leftrightarrow V_2) \right] $$
\beq
\label{19-1e}
+\ T_i^{II} \otimes f_B \Phi_B \otimes f_{V_1} \Phi_{V_1} \otimes f_{V_2} \Phi_{V_2}
\eeq

The first term depends on the form factor $F^{B \to V_1}$ and decay constant $f_{V_2}$, the second is the Annihilation, dependent on the three decay constants $f_B, f_{V_1}, f_{V_2}$.\par

We would like to compare the theoretical prediction for $A_{CP}^{mix}$ for $B_s \to \phi \phi$ with the LHCb measurement. Although other fits at LHCb have been tried that depend on the polarization, the data  (\ref{5-2e},\ref{5-2-1e}) have been obtained assuming the CP-violating phases to be independent of the polarization. To this aim, we will concentrate now on just one polarization, namely the longitudinal one $h = 0$, studied in great detail in \cite{BUCHALLA}.\par

The longitudinal decay amplitude for $B_s \to \phi \phi$ reads, with a sum over the quarks $p = u, c$,
\beq
\label{17bise}
A(\overline{B}_s \to \phi \phi, h = 0) = \sum_{p = u, c}\lambda'_p\ S^{p,0}\ A^0_{\phi \phi} +\ (\lambda'_u + \lambda'_c)\ T^0\ B_{\phi \phi}^0
\eeq

\noi and to get the amplitude $A(B_s \to \phi \phi, h = 0)$ one needs to replace in (\ref{17bise}) $\lambda'_p \to \lambda_p^{'*}\ (p = u, c)$.\par 
For the sake of simplicity, we begin to analize first the trend of the different observables within the QCDF scheme by using central values of the different parameters.
We will later, in sections 5, 6.4 and 7.6, take into account the relevant errors.\par

The amplitude (\ref{17bise}) depends on the quantities listed here below :

$\bullet$ the CKM factors 
\beq
\label{17e}
\lambda'_p = V_{pb}V^*_{ps} \qquad (p = u, c)\ ,
\eeq
\noi given by
\beq
\label{18e}
\lambda '_u =  V_{ub} V^*_{us} = 0.00030 - 0.00076\ i\ ,  \qquad \qquad  \lambda '_c =  V_{cb} V^*_{cs} = 0.039 \ ,
\eeq

$\bullet$ the coefficients $A^0_{\phi \phi}, B^0_{\phi \phi}$ in (\ref{17bise}), that correspond respectively to the direct diagram and to the annihilation diagram, given by
\beq
\label{18-1e}
A^0_{\phi \phi} = i {G_F \over \sqrt{2}}\ m_{B_s}^2 A_0^{B_s \to\phi} (m_{\phi}^2) f_{\phi}\ , \qquad m_{B_s}^2 A_0^{B_s \to\phi} (m_{\phi}^2) f_{\phi} = 3.00\ {\rm GeV^3}
\eeq
\beq
\label{18-2e}
B^0_{\phi \phi} = i {G_F \over \sqrt{2}} f_{B_s} f_{\phi} f_{\phi} \ ,  \qquad f_{B_s} f_{\phi} f_{\phi} = 1.12 \times 10^{-2}\ {\rm GeV^3}
\eeq

\noi for the form factor and decay constants of \cite{BUCHALLA}

$\bullet$ the combinations of coefficients
$$S^{p,0} = 2\left(a_4^{p,0} + a_3^0 +a_5^0 - {1 \over 2} (a_7^{p,0} + a_9^{p,0} + a_{10}^{p,0})\right)$$ 
\beq
\label{17tere}
T^0 = 2\left(b_3^0 + b_4^0 - {1 \over 2}  \left(b_3^{EW,0} + b_4^{EW,0}\right)\right)
\eeq 

\vskip 3 truemm

The coefficients $a_i^0, b_i^0$ in (\ref{17tere}) are given by the short distance Wilson coefficients with the inclusion of the $O(\alpha_s)$ corrections of QCDF that depend now on the longitudinal polarization.\par 

In QCD Factorization, the coefficients $a_i^h$ depend on the helicity. To fix the ideas, we write down their general expression given in the Appendix of \cite{BENEKE-2007},
\beq
\label{20-1e}
a_i^{p,h}(V_1 V_2) = \left(C_i + {C_{i\pm1} \over N_c} \right)N_i^h(V_2)  + {C_{i\pm1} \over N_c} {C_F \alpha_s \over 4 \pi} \left[V_i^h(V_2) + {4 \pi^2 \over N_c} H_i^h(V_1 V_2) \right] + P_i^{p,h}(V_2)
\eeq

\noi where $C_i$ are the short distance Wison coefficients tabulated in Appendix A, the upper (lower) signs correspond to $i$ odd (even), $N_i$ are normalization factors, $V_i$ are the vertex corrections, $H_i$ are the hard scattering corrections, and $P_i$ are the Penguin corrections. The Annihilation corrections $b_i^h$ are not included in this formula.\par

In the limit of disregarding the NLO $\alpha_s$ corrections, the coefficients (\ref{20-1e}) become the usual combinations of short distance Wilson coefficients, e.g.  $a_4^{p,h} \to C_4 + {C_3 \over N_c}$\ ,... where the $C_i$ are given in  Appendix A, while the annihilation coefficients $b_i^h \to 0$.\par

The coefficients at NLO are given in Table 2.\par 
\begin{center}
\begin{tabular}{|c|c|c|}
\hline 
$a_3^0+a_5^0$ & $a_4^{u,0}$ & $a_4^{c,0}$ \\ \hline
$0.002-0.001 i$ & $-0.025-0.016 i$ & $-0.033-0.009 i$ \\ \hline
\end{tabular}

\begin{tabular}{|c|c|c|c|}
\hline 
$(a_7^{u,0}+a_9^{u,0})/ \alpha$ & $(a_7^{c,0}+a_9^{c,0})/ \alpha$ & $a_{10}^{u,0}/ \alpha$ & $a_{10}^{c,0}/ \alpha$ \\ \hline
$-1.84-0.54 i$ & $-1.10-0.02 i$ & $-0.17+0.09 i$ & $-0.17+0.09 i$ \\ \hline
\end{tabular}

\vskip 4 truemm

Table 2. Coefficients at NLO used in QCD Factorization for helicity $h = 0$ \cite{BUCHALLA}.
\end{center}

\noi For Annihilation diagrams one has the parameters of Table 3.

\vskip 3truemm

\begin{center}
\begin{tabular}{|c|c|c|c|c|}
\hline 
Model & $r_A b_3^0$ & $r_A b_4^0$ & $r_A b_3^{EW,0}/\alpha$ & $r_A b_4^{EW,0}/\alpha$ \\ \hline
(1) & 0.003 & -0.003 & -0.035 & 0.013 \\ \hline
(2) & 0.017 - 0.018 \ i & -0.006 + 0.002 \ i & -0.080 + 0.051\ i & 0.023 - 0.009\ i \\ \hline
\end{tabular}
\end{center}

\noi Table 3. Coefficients for $h = 0$ annihilation, with $r_A = B_{\rho \rho}/A_{\rho \rho} = 5. \times 10^{-3}$, from $f_{B_d} = 0.200$ GeV, $f_\rho = 0.209$ GeV, $A_0^{B_d \to \rho}(0) = 0.30$ and $\alpha = 1/129$ \cite{BUCHALLA}. Model (1) uses the default value $X_A = \ln {m_B \over \Lambda_h}$ for the power-suppressed annihilation contributions \cite{BUCHALLA}, and the results for model (2) are obtained in Appendix C, with $X_A = \left(1 + \rho_A e^{i \phi_A} \right) \ln {m_B \over \Lambda_h}$ and $\rho_A = 0.6, \phi_A = -40^0$ from \cite{BENEKE-2007}.
\vskip 3truemm

From the parameters of Tables 2 and 3 we find the factors relevant to eq. (\ref{17bise}),
$$S^{u,0} = a_4^{u,0} + a_3^0 +a_5^0 - {1 \over 2} (a_7^{u,0} + a_9^{u,0} + a_{10}^{u,0}) = -0.0304 - 0.0306 i$$
$$S^{c,0} = a_4^{c,0} + a_3^0+a_5^0 - {1 \over 2} (a_7^{c,0} + a_9^{c,0} + a_{10}^{c,0}) = -0.0522 - 0.0205 i$$
\beq
\label{22-1e}
T^0 = b_3^0 + b_4^0 - {1 \over 2}  \left(b_3^{EW,0} + b_4^{EW,0} \right) = 0.0342
\eeq

\noi We have, with a sum over $p = u, c$, the ratio
\beq
\left({A(\overline{B}_s \to \phi \phi, h = 0) \over A(B_s \to \phi \phi, h = 0)} \right)_{QCDF} = - {\sum_{p=u,c} \lambda'_p\ S^{p,0}\ A^0_{\phi \phi} +\ (\lambda'_u + \lambda'_c)\ T^0\ B_{\phi \phi}^0 \over \sum_{p=u,c} (\lambda'_p)^*\ S^{p,0}\ A^0_{\phi \phi} +\ [(\lambda'_u)^* + (\lambda'_c)^*]\ T^0\ B_{\phi \phi}^0}
\label{22-2-1}
\eeq

\noi Using the central values given by Tables 2 and 3 (Model (1)), and (\ref{18-1e},\ref{18-2e}) we get
\beq
\label{22-3e}
\left({A(\overline{B}_s \to \phi \phi, h = 0) \over A(B_s \to \phi \phi, h = 0)}\right)_{QCDF} = -1.003 + 0.031 i 
\eeq 

\noi Therefore, the departure from the NF value $- {V_{tb}V_{ts}^* \over V_{tb}^*V_{ts}} = -0.999 + 0.035 i$ is of the order of the percent.\par 

\subsection{Time-dependent CP violation in QCD Factorization}

We will now use the notation of Bartsch et al. \cite{BUCHALLA}, and their formulas (74) and (113),
\beq
\label{5-2-4bise} 
a_p = S^{p,0}\ m_{B_s}^2 A_0^{B_s \to\phi} (m_{\phi}^2) f_{\phi} + T^0\ f_{B_s} f_{\phi} f_{\phi}
\eeq

\noi where $a_p (p = u, c)$ are the coefficients of $i {G_F \over \sqrt{2}} \lambda'_p  \ (p = u, c)$ of the total $\overline{B}_s \to \phi_L \phi_L$ amplitude (\ref{17bise}), with the CKM factors $\lambda'_p  \ (p = u, c)$ defined by (\ref{17e}). \par
Let us write the Standard Model quantity 
\beq
\label{5-2bise} 
\lambda^{SM}_{\phi \phi} = \left( {q \over p} \right)_{B_s} \left({A(\overline{B}_s \to \phi \phi, h = 0) \over A(B_s \to \phi \phi, h = 0)} \right)_{QCDF}
\eeq

\noi with the mixing computed from the box diagram
\beq
\label{5-3bise} 
\left({q\over p}\right)_{B_s} \simeq - {V^*_{tb} V_{ts} \over V_{tb} V^*_{ts}} \simeq - {V_{ub}^*V_{us} + V_{cb}^*V_{cs} \over V_{ub}V_{us}^* + V_{cb}V_{cs}^*} = - {\lambda^{'*}_u + \lambda^{'*}_c \over \lambda'_u + \lambda'_c}
\eeq

\noi and the decay amplitude estimated using QCDF,
\beq
\label{5-5bise} 
\left({A(\overline{B}_s \to \phi \phi, h = 0) \over A(B_s \to \phi \phi, h = 0)} \right)_{QCDF} 
= - {\lambda'_c a_c + \lambda'_u a_u\over \lambda^{'*}_c a_c +\lambda^{'*}_u a_u} = - {a_c + \lambda^2 (\rho-i\eta) a_u\over a_c + \lambda^2 (\rho+i\eta) a_u}
\eeq

\noi where $\lambda = 0.226$, $\rho = 0.141$ and $\eta = 0.357$ are CKM parameters. 

The time-dependent CP asymmetry writes, in the limit of neglecting the width difference $\Delta \Gamma_s$,
\beq
\label{5-7bise} 
A^{CP}_\phi (t) = S_\phi\ {\rm sin} (\Delta M_s t) - C_\phi\ {\rm cos}(\Delta M_s t)
\eeq
\beq
\label{5-8bise} 
S_\phi = { 2 {\rm Im} \lambda_{\phi \phi}  \over 1 + \mid \lambda_{\phi \phi}\mid^2}\ , \ \ \ C_\phi = {1 - \mid\lambda_{\phi \phi}\mid^2  \over 1 + \mid \lambda_{\phi \phi}\mid^2}\ , \ \ \ \lambda_{\phi \phi} = -{M_{12}^* \over \mid M_{12} \mid} {A(\overline{B}_s \to \phi \phi, h = 0) \over A(B_s \to \phi \phi, h = 0)}
\eeq 

\vskip 3 truemm

\noi Particularizing to the longitudinal polarization $\phi_L \phi_L$ and QCDF, and expanding in powers of the Wolfenstein parameter $\lambda$ one finds,
$$\lambda^L_{\phi \phi} = \left( {q \over p} \right)_{B_s} \left( {A(\overline{B}_s \to \phi \phi, h = 0) \over A(B_s \to \phi \phi, h = 0)}\right)_{QCDF} = {\lambda^{'*}_u + \lambda^{'*}_c \over \lambda'_u + \lambda'_c} \ {\lambda'_c a_c + \lambda'_u a_u\over \lambda^{'*}_c a_c + \lambda^{'*}_u a_u}$$
\beq
\label{5-9bise} 
\simeq 1 + 2 \lambda^2 \eta\ \left[-{\rm Im} \left({a_c-a_u \over a_c}\right) + i\ {\rm Re}\left({a_c-a_u \over a_c}\right)\right]
\eeq

\noi In the limit of NF one has $a_c = a_u$ and therefore $\lambda^{NF}_{\phi \phi} = 1$.

\noi One gets, with the definition (\ref{5-1-2e}),
\beq
\label{5-9-1e} 
\mid \lambda_{\phi \phi}^L \mid\ \simeq 1 - 2 \lambda^2 \eta\ {\rm Im} \left( {a_c-a_u \over a_c} \right)
\eeq
\beq
\label{5-9-2e} 
\phi_{\phi \phi}^L \simeq - 2 \lambda^2 \eta\ {\rm Re} \left( {a_c-a_u \over a_c} \right)
\eeq

From the values of the coefficients given in Tables 2 and 3 for Model (1) and the form factor $A^0_{\phi \phi} = 0.47$ \cite{BUCHALLA}, one gets 
$$a_c = -0.156 - 0.062 i \ , \ \ \ a_u = -0.091 - 0.091 i \ , \ \ \  {a_c - a_u \over a_c} = 0.296 - 0.308 i$$
\beq
\label{5-10bise} 
\mid a_c \mid\ = 0.168\ {\rm GeV^3} \ , \qquad \mid a_c - a_u \mid\ = 0.072\ {\rm GeV^3}
\eeq

\noi and therefore
\beq
\label{5-11-1e} 
\mid \lambda_{\phi \phi}^L \mid\ \simeq 1.011\ , \qquad \qquad \phi_{\phi \phi}^L \simeq - 0.011
\eeq

\noi hence,
\beq
\label{5-11tere} 
S_{\phi_L} = 2 \lambda^2 \eta\ {\rm Re} \left({a_c  -a_u \over a_c} \right) = 0.011\ , \qquad C_{\phi_L} = 2 \lambda^2 \eta\ {\rm Im} \left({a_c  -a_u \over a_c} \right) = -0.011
\eeq

On the other hand, from the values of the coefficients given in Tables 2 and 3 for Model (2) and the form factor $A^0_{\phi \phi} = 0.38$ \cite{BENEKE-2007}, that we will adopt below, one gets 
\beq
\label{5-11-4e} 
a_c = -0.076 - 0.122 i \ , \ \ \ a_u = -0.023 - 0.147 i
\eeq

\noi and therefore
\beq
\label{5-11-1bise} 
\mid \lambda_{\phi \phi}^L \mid\ \simeq 1.015\ , \qquad \qquad \phi_{\phi \phi}^L \simeq 10^{-3}
\eeq

\subsection{Bounds on the modulus and phase of $\lambda_{\phi \phi}^L$} 

Going beyond the central values of Tables 2 and 3, that lead to (\ref{5-11-1e}), following Bartsch et al. \cite{BUCHALLA} one can bound the modulus and phase of $\lambda_{\phi \phi}^L$.\par 
Bartsch et al. have computed  $\mid a_c - a_u \mid$ in QCD Factorization, finding
\beq
\label{5-12bise}  
\mid a_c - a_u \mid\ \simeq 0.057\ {\rm GeV^3}
\eeq 
On the other hand, since $\mid \lambda'_u \mid\ <<\ \mid \lambda'_c \mid$, one can compute $\mid a_c \mid$ from the longitudinal branching ratio $B(\overline{B}_s \to \phi \phi, h = 0)$,
\beq
\label{5-13bise}
\mid a_c \mid\ = 0.177\ {\rm GeV^3} \left[ {B(\overline{B}_s \to \phi \phi, h = 0) \over 18.7 \times 10^{-6}} \right]^{1/2} \left[ {1.515\ {\rm ps} \over \tau_{B_s}} \right]^{1/2}
\eeq

\noi At LHCb \cite{LHCbPHIPHI} the longitudinal fraction has been measured, as shown in Table 1, yielding,
\beq
\label{5-15bise}
B(\overline{B}_s \to \phi \phi, h = 0) \simeq 0.38 \times  B(\overline{B}_s \to \phi \phi) \simeq 7.1 \times 10^{-6}
\eeq

\noi and therefore,
\beq
\label{5-16bise}
\mid a_c \mid\ \simeq 0.11
\eeq

On the other hand, using (\ref{5-12bise},\ref{5-16bise}), one has the upper bounds \cite{BUCHALLA}, 
$$
2 \lambda^2 \eta\ {\rm Re} \left({a_c-a_u \over a_c}\right) \leq 2 \lambda^2 \eta\ {\mid a_c-a_u \mid \over \mid a_c \mid} \simeq 0.019$$
\beq
2 \lambda^2 \eta\ {\rm Im} \left({a_c-a_u \over a_c}\right) \leq 2 \lambda^2 \eta\ {\mid a_c-a_u \mid \over \mid a_c \mid} \simeq 0.019
\label{6-13tere}
\eeq

\noi and therefore,
\beq
\label{5-10-1e} 
\mid \lambda_{\phi \phi}^L \mid\ \geq 0.981
\eeq
\beq
\label{5-10-2e} 
\phi_{\phi \phi}^L \geq -0.019
\eeq

\noi These bounds are not very constraining, and lie within the range of LHCb (\ref{5-2-1e}).

Following Bartsch et al. \cite{BUCHALLA} one finds the upper bound,
\beq
\label{5-17bise}
S_\phi \leq 2 \lambda^2 \eta {\mid a_c - a_u \mid \over \mid a_c \mid} \simeq 0.019
\eeq

\noi and similarly for the modulus $\mid C_\phi\mid$.

\section{Comparison of LHCb data with the SM and QCD Factorization}

To summarize Section 3.2, let us recall that we have found, with top dominance in mixing and NF in decay,
\beq
\mid \lambda_{\phi \phi}\mid \ = 1\ ,  \qquad \qquad \phi_{\phi \phi} = 0
\label{4-2e}
\eeq

The different effects that go beyond NF, given by QCDF (\ref{5-11-1e}), obtained from the central values of the QCDF coefficients of Tables 2 and 3, are consistent within errors with the LHCb polarization-independent fit (\ref{5-2-1e}).\par
One could conversely look for the range of the QCDF parameters that are allowed by the LHCb data.\par
From (\ref{5-9bise}) we get the experimental constraint
\beq
1 + 2 \lambda^2 \eta\ \left[-{\rm Im} \left({a_c-a_u \over a_c}\right) + i\ {\rm Re}\left({a_c-a_u \over a_c}\right)\right] = \lambda^{LHCb}_{\phi \phi}
\label{6-1bise}
\eeq

\noi From the LHCb data for $\lambda^{LHCb}_{\phi \phi}$ and the SM value $\lambda^{SM}_{\phi \phi}$, we realize that they are rather close and we can expand the precedent relation in the small quantities 
\beq
\lambda^2, \qquad \mid\lambda^{LHCb}_{\phi \phi}\mid - 1, \qquad \phi_{\phi \phi}^{LHCb}
\label{6-3bise}
\eeq

\noi giving the two relations
\beq
\mid\lambda^{LHCb}_{\phi \phi}\mid - 1 = -2 \lambda^2 \eta\ {\rm Im} \left({a_c-a_u \over a_c}\right)
\label{6-5bise}
\eeq
\beq
\phi_{\phi \phi}^{LHCb} = -2 \lambda^2 \eta\ {\rm Re}\left({a_c-a_u \over a_c}\right)
\label{6-6bise}
\eeq

From the values of LHCb (\ref{5-2-1e}), one gets
\beq 
2 \lambda^2 \eta\ {\rm Im} \left({a_c-a_u \over a_c}\right) = 0.01 \pm 0.05
\label{6-10bise}
\eeq
\beq
2 \lambda^2 \eta\  {\rm Re}\left({a_c-a_u \over a_c}\right) = 0.073 \pm 0.115
\label{6-11bise}
\eeq

\noi to be compared with (\ref{5-11tere}), obtained from the central values in Tables 2 and 3(1).

Interestingly, from the LHCb data, or in the future from other experiments, if one assumes the CKM matrix for the electroweak sector, one could measure the QCDF quantities ${\rm Re} \left({a_c-a_u \over a_c}\right)$ and $ {\rm Im} \left({a_c-a_u \over a_c}\right)$.

\subsection{Including New Physics in mixing}

We now go beyond the SM, and compare theory and experiment including NP contributions.
A usual parametrization of a NP contribution to mixing is the following (see ref. \cite{CKMfitter} and references therein), 
$$M_{12}^s = M_{12}^{SM,s} \Delta_s = M_{12}^{SM,s} \left( {\rm Re}\ \Delta_s + i {\rm Im}\ \Delta_s \right) =\ M_{12}^{SM,s} \mid \Delta_s \mid e^{i \phi_s^\Delta}$$
\beq
\phi_s = \phi_s^{SM} + \phi_s^\Delta
\label{5-1-3e}
\eeq

\noi where $M_{12}^{SM,s}, \phi_s^{SM}$ stand for Standard Model values and the complex parameter $\Delta_s$ for New Physics, the SM corresponding to $({\rm Re}\Delta_s, {\rm Im}\Delta_s) = (1,0)$. On the other hand, we observe that the LHCb measurement is rather close to the SM values, but we would like to know which domain in the plane $({\rm Re}\Delta_s, {\rm Im}\Delta_s)$ is still allowed from the data.\par

Including NP contributions, as in (\ref{5-1-3e}), one has from LHCb the experimental constraint on the NP parameters,
\beq
\lambda^{SM}_{\phi \phi} \sqrt{{\Delta_s^* \over \Delta_s}} = \lambda^{LHCb}_{\phi \phi} 
\label{5-1-4e}
\eeq

\noi where $\lambda^{SM}_{\phi \phi}$ is given by (\ref{5-9bise}) and $\lambda^{LHCb}_{\phi \phi}$ by (\ref{5-2e},\ref{5-2-1e}). 

Let us now look for the constraint on the NP mixing parameters.
From (\ref{5-1-3e},\ref{5-1-4e}) and the expansion (\ref{5-9bise}) we get
\beq
\left\{1 + 2 \lambda^2 \eta\ \left[-{\rm Im} \left({a_c-a_u \over a_c}\right) + i\ {\rm Re}\left({a_c-a_u \over a_c}\right)\right]\right\} \sqrt{{{\rm Re}\Delta_s - i{\rm Im}\Delta_s \over {\rm Re}\Delta_s + i{\rm Im}\Delta_s}} = \lambda^{LHCb}_{\phi \phi}
\label{6-1bise}
\eeq

\noi or, up to small SM terms of $O(\lambda^4)$, 
\beq
\left[1 - 2 \lambda^2 \eta\ {\rm Im} \left({a_c-a_u \over a_c}\right) \right] e^{-i\phi_{\phi \phi}^{SM}} e^{-i\phi_s^\Delta} = \lambda^{LHCb}_{\phi \phi}
\label{6-1tere}
\eeq

\noi with (\ref{5-9-2e}),
\beq
\phi_{\phi \phi}^{SM} = - 2 \lambda^2 \eta\ {\rm Re} \left({a_c-a_u \over a_c}\right)  
\label{6-1-4e}
\eeq

\noi and, from (\ref{5-2-1e}), eqn. (\ref{6-1tere}) implies the two constraints,
\beq
1 - 2 \lambda^2 \eta\ {\rm Im} \left({a_c-a_u \over a_c}\right) =\ \mid \lambda^{LHCb}_{\phi \phi} \mid\ = 0.99 \pm 0.05 
\label{6-1-5e}
\eeq
\beq
\phi_s^\Delta = - \phi_{\phi \phi}^{SM} + \phi_{\phi \phi}^{LHCb} = - \phi_{\phi \phi}^{SM} + (-0.073 \pm 0.115)
\label{6-1-6e}
\eeq

\noi Notice that the first constraint (\ref{6-1-5e}) concerns only the SM quantity  $2 \lambda^2 \eta\ {\rm Im} \left({a_c-a_u \over a_c}\right) = -0.011$ (\ref{5-11tere}), while the second constraint (\ref{6-1-6e}) involves SM model and NP quantities.\par
The first constraint (\ref{6-1-5e}) writes 1.011 in the l.h.s. vs. the r.h.s. LHCb value $0.99 \pm 0.05$, and it is satisfied within errors.

From the second constraint (\ref{6-1-6e}), and (\ref{5-11-1e}) $\phi_{\phi \phi}^{SM} \simeq -0.011$, one obtains for the NP parameter
\beq
{{\rm Im}\Delta_s \over {\rm Re}\Delta_s} \simeq -0.062 \pm 0.115
\label{6-14bise}
\eeq

\noi i.e.,
\beq
-0.177 \leq {{\rm Im}\Delta_s \over {\rm Re}\Delta_s} \leq 0.053
\label{6-17bise} 
\eeq

\noi that gives the domain in Fig. 4, from the $1 \sigma$ error of the LHCb CP phase. 
\begin{center}
\includegraphics[scale=1.]{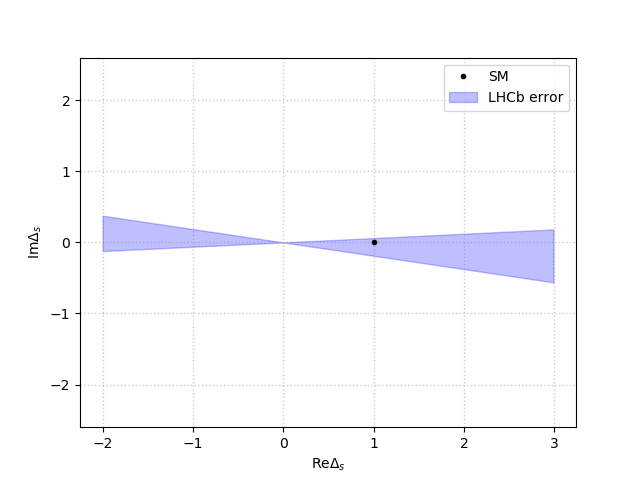}
\end{center}
Fig. 4. Domain allowed in the plane Re$\Delta_s$-Im$\Delta_s$ from $B_s \to \phi \phi$ LHCb data. If one takes into account the constraints from other observables, as exposed for example in ref. \cite{CKMfitter}, then only the domain for ${ \rm Re} \Delta_s > 0$ remains.

\vskip 5 truemm

This domain is comparable to the domain allowed by other observables, in particular from the decay $\overline{B}_s \to J/\psi \phi$, e.g. by CKMfitter \cite{CKMfitter}.\par
We conclude that the mode $B_s \to \phi \phi$, combined with other observables, can be useful in the future to look for or to discard NP contributions.\par

\section{Expected sensitivity at FCCee and implications in the search for NP}

\noindent FCCee is a fantastic source of B mesons when operating at the Z-pole thanks to the relatively large cross section for the production of Z bosons ($\sim 42.9$ nb) and the large instantaneous luminosity of $2.3\ 10^{36}$\ cm$^{-2}$s$^{-1}$, which is planned for. Futhermore it is foreseen to operate the collider with at least 2 detectors for 4 years at the Z-pole so that a total integrated luminosity of 150 ab$^{-1}$ can be accumulated. As a consequence, a large number of $\overline{B_s}(B_s) \to \phi\phi$ ($\sim 9.4\ 10^5$) is expected as summarized in Table 4. In the following, we use a parametrized detector to evaluate the sensitivities, which can be achieved at FCCee~\cite{fccee:1,fccee:2,fccee:3}.
$$ \begin{tabular}{cccc}
\hline
&  & $\displaystyle {\mathrm {E_{cm} = 91.2\ GeV }}$   &  \\
&  & $\displaystyle {\mathrm {\int L = 150 ab^{-1}}}$   &  \\

$\displaystyle {\mathrm {\sigma (e^+e^- \to Z )}} $ &
$\displaystyle {\mathrm {number}} $ &
$\displaystyle {\mathrm {}} $ &
$\displaystyle {\mathrm {Number \ of}} $ \\

$\displaystyle {\mathrm {nb}} $ & 
$\displaystyle {\mathrm {of \ Z} } $ &
$\displaystyle {\mathrm{f[Z\to \overline{B_s}(\overline{B_d})]}} $ &
$\displaystyle {\mathrm{produced \ \overline{B_s}(\overline{B_d}) }} $\\
\hline \hline \\

$\displaystyle \sim 42.9$ &
$\displaystyle {\mathrm {\sim 6.4\ 10^{12}}}$ &
$\displaystyle {\mathrm {0.0159 (0.0608)}} $ &
$\displaystyle \sim 1\ 10^{11}(3.9\ 10^{11})$\\ 
\hline
& & & \\

$\displaystyle {\mathrm {\overline{B}_{d,s}\ decay}} $ &
$\displaystyle {\mathrm {\phi \ decay}} $ & 
$\displaystyle {\mathrm {Final}} $ &
$\displaystyle {\mathrm {Number \ of}} $ \\

$\displaystyle {\mathrm {Mode}}  $ &
$\displaystyle {\mathrm {Mode} } $ &
$\displaystyle {\mathrm{State}} $ &
$\displaystyle {\mathrm{\overline{B}_{d,s} \ decays}} $ \\ 
\hline \hline
& & & \\

$\displaystyle \overline{B_d} \to \phi\phi$ &
$\displaystyle {\mathrm {\phi \to K^+K^-}}$ &
$\displaystyle {\mathrm {K^+K^-K^+K^-}} $ &
$\displaystyle \sim2.9\ 10^2 $\\ 

$\displaystyle \overline{B_s} \to \phi\phi$ &
$\displaystyle {\mathrm {\phi \to K^+K^-}}$ &
$\displaystyle {\mathrm {K^+K^-K^+K^-}} $ &
$\displaystyle \sim 4.7\ 10^5 $\\ 


\hline

\end{tabular}   $$

\noi Table 4. The expected numbers of produced $\overline{B_s}(\overline{B_d})$ decays to the specific decay mode $\overline{B_s}(\overline{B_d}) \to \phi\phi$ at FCC-ee at a center of mass energy of 91.2 GeV over 4 years with 2 detectors. These numbers have to be multiplied by 2 when including $B_s(B_d)$ decays. The branching fractions of the PDG~\cite{pdg:1} have been used for $\overline{B_s} \to \phi\phi$, while it is estimated to $\sim 3\ 10^{-9}$ using QCDF for $\overline{B_d} \to \phi\phi$.\par 

\subsection{Generic detector resolutions}

\noindent In order to carry out experimental studies, we define a generic detector, the resolution of which is  parametrized as follow  :
\begin{equation}
\begin{array}{lccl} 
\mathrm{Acceptance :}& |\cos \theta|&<&0.95\\
\hline
\mathrm{Charged \ particles :} & \\
\mathrm{ p_T\ resolution :}& {\sigma (p_T) \over p_T^2}  & = & 2. \times 10^{-5} \ \oplus \ {1.2 \times 10^{-3}\over p_T \sin \theta}\\
\mathrm{ \phi , \theta \ resolution :}& \mathrm{\sigma (\phi , \theta) \ \mu rad } & = &  18 \ \oplus \ {1.5 \times 10^{3} \over p_T\sqrt[3]{\sin \theta} }\\
\mathrm{Vertex \ resolution :}& \mathrm{\sigma (d_{Im}) \ \mu m} & = &  1.8 \ \oplus \ {5.4 \times 10^{1} \over p_T\sqrt{\sin \theta} }\\
\hline
\mathrm{e,\gamma \ particles :} & \\
\mathrm{Energy\ resolution:}&{\sigma (E) \over E}  & = & {5 \times 10^{-2} \over \sqrt{E}} \ \oplus \ 5 \times 10^{-3}\\
\mathrm{EM \ \phi , \theta \ resolution :}& \mathrm{\sigma (\phi , \theta) \ m rad } & = &  {7 \over \sqrt{E} }\\
\hline \hline
\end{array}
\label{eq:track_resolution}
\end{equation}

\vskip 3 truemm

\noindent where $\theta, \phi $ are the particles' polar and azymutal angles respectively, $p_T$ (in GeV) the track transverse momentum, $E$ the $e^\pm\ ,\gamma$ energy and $\mathrm{d_{Im}}$ the tracks' impact parameter.

\vskip 10pt

\noindent With the detector resolutions in (\ref{eq:track_resolution}) and a reasonably good particle identification system (ToF + cluster counting for $dE/dx$), a clean signal for $\overline{B_s}(B_s) \to \phi\phi$ is expected.  The main source of background is the combinatorial one, which is however expected to be small (Background/Signal $< 10\%$), thanks to a good particle identification system and the excellent mass resolution for $\phi$ and $B_s$ as shown in Figure 5 for the latter.

\includegraphics[scale=0.9]{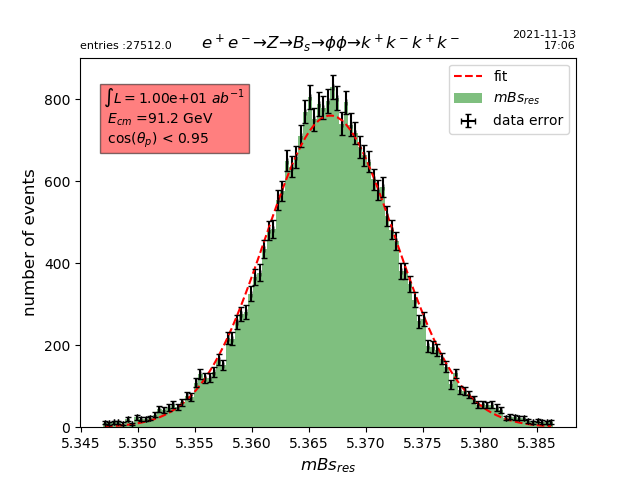}

Fig. 5. $B_s$ mass resolution for $\overline{B_s} (B_s) \to \phi\phi \to K^+K^-K^+K^- $ decay. One obtains $\sigma(m_{B_s}) \simeq 5.7$ MeV. The geometric acceptance of the detector ($|\cos \theta |< .95$) leads to an efficiency of 86$\%$ for this 4-body final state. The combinatorial background is not shown in this figure.

\subsection{Experimental expectations for CP violation parameters at FCCee}

We use eq. (\ref{5-3tere}) to obtain the decay widths for $\overline{B}_s(B_s) \to \phi \phi$ and we include the wrong tagging fraction $\omega$,

$$\Gamma (\overline{B}_s \to \phi\phi) = |< \phi\phi |B_s >|^2 \times e^{-\Gamma t} \{ \cosh {\Delta\Gamma\over 2} - (1-2\omega ) A^{dir}_{CP}\cos \Delta m t$$
$$+\ A_{\Delta\Gamma} \sinh {\Delta\Gamma\over 2} - (1-2\omega ) A^{mix}_{CP} \sin \Delta m t \}$$
$$\Gamma (B_s \to \phi\phi ) = | < \phi\phi |B_s >|^2\times   e^{-\Gamma t}  \{ \cosh {\Delta\Gamma\over 2} + (1-2\omega ) A^{dir}_{CP}\cos \Delta m t$$
\beq
\label{eq:Bs_CPdecay}
+\ A_{\Delta\Gamma} \sinh {\Delta\Gamma\over 2} + (1-2\omega )  A^{mix}_{CP} \sin \Delta m t \}
\eeq

It is important to note that the  $\Delta\Gamma -$dependent terms enable one to solve the ambiguity for the determination of the CP violating phase $\phi_{\phi\phi}$. We have verified that keeping $\Delta\Gamma$ as a free parameter in the fit does not degrade significatively the sensitivity obtained in this study for the parameters $|\lambda_{\phi\phi}|$ and the CP violating phase $\phi_{\phi\phi}$ . However the determination of $\Delta\Gamma $ is not the main topic in this study as it can be obtained with higher precision by studying other modes such as $B_s\to J/\psi \ \phi$ and $B_s\to J/\psi \ \eta$.

As shown in equation (\ref{5-5e}),  $A^{mix}_{CP}$ depends on $\eta_f$, the CP eigenvalue of the final state. For a V$_1$-V$_2$ final state such as $\phi \phi$, $\eta_f$ varies as a function of  the polarization state. For $\phi \phi$, it is $\eta_f=+1$ for the longitudinal and parallel polarizations while $\eta_f=-1$ for the perpendicular polarization. In case one does not disentangle the polarization by doing a full angular analysis, one would have an effective $\eta_{\phi \phi}^{eff} =1-2f_\perp = 0.416\pm 0.018$ as obtained from the experimentally measured polarizations~\cite{01111} summarized in Table 1. Since it is expected in the Standard Model that $|\lambda_{\phi\phi}|\simeq 1$, we simplify $A^{mix}_{CP} \simeq - \eta_{\phi \phi}^{eff} \sin\phi_{\phi\phi}$, where $\phi_{\phi\phi}$ is the phase responsible of CP violation.

\vskip 10pt
\noindent As one can see in equation (\ref{eq:Bs_CPdecay}), CP violating effects are damped by the fraction of wrong tagging, $\omega$ (Table 5 shows typical tagging performances of some experiments). 

\vskip 3 truemm

$$\begin{tabular}{cccc}

\hline
Tagging Merit & LEP & BaBar & LHCb  \\
\hline\hline
$\epsilon(1-2\omega)^2$ & 25-30\% & ~30\% & ~6\%\\
\hline

\end{tabular}$$

\vskip 2 truemm

Table 5. Typical tagging Figure of Merit for some experiments. $\epsilon$ is the tagging efficiency and $\omega$, the wrong tagging fraction, which is in the range $0-0.5$.

\vskip 5 truemm

It is thus essential to measure this factor precisely. Fortunately, the large statistics at FCCee enables one to measure the value of $\omega$ very precisely, for example with the decay $\overline{B_s}\to D^+_s \pi^-\to \phi \pi^+\pi^-$, which is similar to the  final state $\phi\phi$ (see \cite{ALEKSAN} for details). With this very clean and abundant mode, $\omega$ can be determined with a negligible error. In the following, we assume $\omega = 0.25$, which is rather conservative since it was achieved at LEP. 

\vskip 10pt
Another important factor to consider is the detector resolution for the reconstruction of the $B_s$ vertex. A detailed simulation study has been done in a previous paper for the mode $B_s \to J/\psi \phi$ \cite{ALEKSAN}, which is a final state rather similar to $\phi\phi$. We have found that a resolution of $\sim 20\ \mu m$ can be obtained, which does not affect significatively the resolution on $\phi_{\phi\phi}$ (i.e. $2-3\%$ level) , and can thus be safely corrected for. Indeed, this vertex resolution of $20\ \mu m$ has to be compared to the average $B_s$ flight distance of about 3mm at the Z-pole.\par

\includegraphics[scale=0.9]{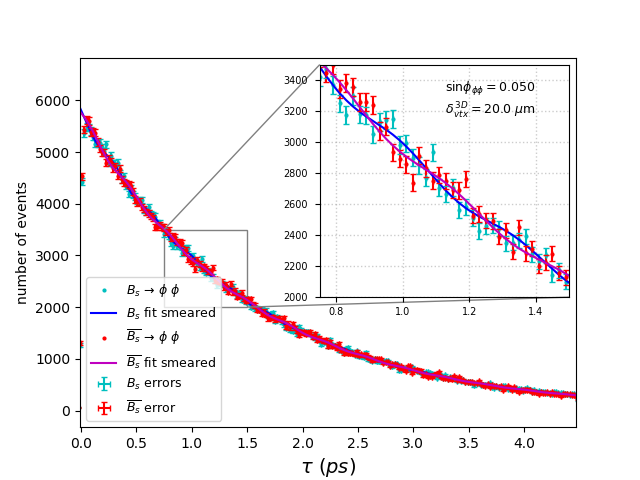}	

Fig. 6. The time-dependent distributions for the simulated $\overline{B_s} (B_s)\to \phi\phi$ events are shown with the statistics of 150 ab$^{-1}$ expected at FCCee.
The input parameters $|\lambda_{\phi\phi}|=1$, $\phi_{\phi\phi}=0.05$, $\eta_{\phi\phi} = 0.416$, $\Delta \Gamma = 0$, and $\omega = 0.25$ have been used.

\vskip 3truemm

To estimate the sensitivity, which can be achieved at FCCee, we assume the input parameters $|\lambda_{\phi\phi}|=1$, $\phi_{\phi\phi}=0.05$ and $\eta_{\phi\phi}^{eff} = 0.416$, and generate the time-dependent signal events accordingly. Figure 6 shows a typical time-dependent spectrum of the signal, which is expected in an experiment with the integrated statistics of 150 ab$^{-1}$.

We then repeat such experiment 5000 times and extract the distributions for the fitted variables $|\lambda_{\phi\phi}|$ and $\phi_{\phi\phi}$. From these gaussian distributions, we derive the sensitivities and obtain :
\begin{equation}
\begin{array}{ccccccc} 
\delta(|\lambda_{\phi\phi}|) &  = & 0.004 & \qquad \qquad \mathrm{    and    } \qquad \qquad &\delta(\phi_{\phi\phi}) &  = & 0.009 \mathrm{\ rad}\\
\end{array}
\label{eq:Bs_CPsens}
\end{equation} 

\noindent Interestingly, should the value of $\phi_{\phi\phi}$ be as large as 0.05, one would be able to exclude $\phi_{\phi\phi}=0$ at the  5$\sigma$ level, which would thus invalidate the Naive Factorization as well as the QCD Factorization, and would point towards physics beyond the SM.\par

We show in Fig. 7 the $1\sigma$ and $2\sigma$ contours in the plane $\phi_{\phi \phi}$ versus $\mid \lambda_{\phi \phi} \mid$ for the expected sensitivity at FCCee. Although it still remains to be studied in detail, we believe that the presence of the combinatorial background at the $10\%$ level would not degrade these resolutions noticeably, thanks to the high statistics allowing one to study the background very precisely.\par

As mentioned, the overall fit enables also the determination of $\Delta \Gamma$. The sensitivity is :
\beq
\delta(\Delta \Gamma) = 0.004 \times 10^{12} s^{-1}
\label{DeltaGamma}
\eeq

\includegraphics[scale=0.9]{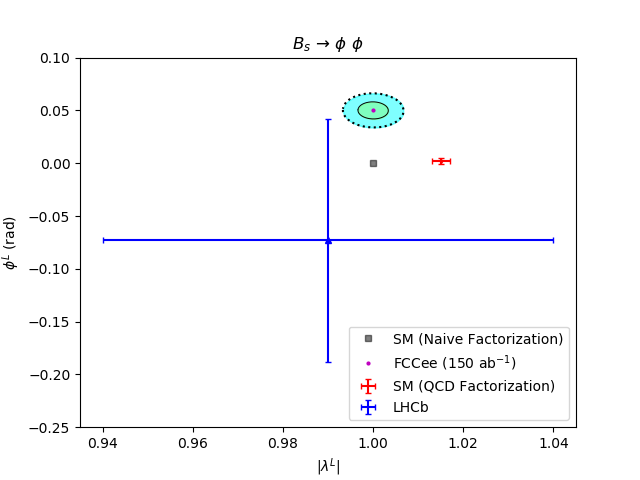}

Fig. 7. Typical contour plot for the expected sensitivity on $\phi_{\phi\phi}$ versus $|\lambda_{\phi\phi}|$ at FCCee. The solid (dotted) ellipse curve is for 1$\sigma$ (2$\sigma$).
The input parameters are $|\lambda_{\phi\phi}|=1$, $\delta(|\lambda_{\phi\phi}|) = 0.005$, $\phi_{\phi\phi}=0.05$, $\delta(\phi_{\phi\phi}) = 0.01$, $\eta_{\phi\phi}^{eff} = 1 - 2 f_\perp = 0.416$ and wrong tagging fraction $\omega = 0.25$. Are also shown the present sensitivity at LHCb and the expected SM theoretical values with NF or QCDF (\ref{5-11-1bise}).

\vskip 4 truemm

Finaly, we wish to stress that the sensitivities in equation (\ref{eq:Bs_CPsens}), in particular for $\phi_{\phi\phi}$ are not optimal as we did not carry out a full angular analysis. With such an angular analysis, one can expect to reduce further this error. Although one has to keep this possibility in mind, we do not assume such an improvement in the following of this paper, to remain on the conservative side. The study of this mode is particularly interesting for searching Beyond the SM physics and thus higher integrated luminosity would be most welcome either with higher instantaneous luminosity or enabling four interaction regions, or ideally both !

\subsection{Contraints on FCC detectors}

\noindent We explore the constraints relevant for this study that are to be considered when designing the detectors. To this end, we detail in Table 6 how the resolutions scale for the considered parameters.\par
Let us be more explicit on the experimental constraints.

$$ \begin{tabular}{cclc}
\hline 

$\displaystyle {\mathrm {Parameters}} $ &
$\displaystyle {\mathrm {Errors \ scaling}} $ \\

\hline \hline \\

$\displaystyle \phi_{\phi\phi}$ &
$\displaystyle {\mathrm {\frac{1}{ \eta_{\phi\phi}(1-2\omega)\sqrt{N} }}}$  \\ 

$\displaystyle |\lambda_{\phi\phi}|$ &
$\displaystyle {\mathrm {\frac{1}{(1-2\omega)\sqrt{N}}}}$ \\

$\displaystyle \Delta\Gamma$ &
$\displaystyle {\mathrm {\frac{1}{\eta_{\phi\phi}\sqrt{N}}}}$  \\

\\

\hline

\end{tabular}   $$

\vskip 4 truemm

Table 6. The scaling of the errors for the various parameters determining the shape of the time dependent distributions for the decay mode $B_s(\overline{B}_s) \to \phi\phi$.

\subsubsection{$\sqrt{N}$ : Increasing the statistics}
Several aspects are to be considered to increase the statistics :
\begin{itemize}
\item Increasing the instantaneous luminosity of the accelerator.
\item Increasing the integrated luminosity, for example by operating 4 detectors and/or running more time at the Z-pole.
\item Increasing the acceptance of the detector and the reconstruction efficiency. For the latter, one has to investigate the benefit of a large detection volume combined with a larger number of tracking points. This is particularly important since the K$^+$K$^-$  tracks issued from each $\phi$ meson are very close to each other. A gazeous tracking detector (TPC or wire chambers) might be advantageous.
\item An additional requirement concerns the overall tracking system resolution. Indeed, with an excellent momentum resolution, one is able to reduce the invariant $B_s-$mass resolution and hence reduce considerably the background. Not only point resolution is important but, maybe even more, very low material budget of the tracking system is of prime importance. Here again a gazeous tracking volume may be advantageous.
\end{itemize}

\subsubsection{$(1-2\omega)$ : decreasing the fraction of wrong tagging}
In the present study, we have assumed conservatively $\omega=25\%$ as it was achieved at LEP. Two main characteristics are important for reducing further the wrong tagging fraction:
\begin{itemize}
\item Excellent vertex resolution to identify the secondary and tertiary vertices, which are typical for the $B$ decays. Such analysis is also mandatory to measure precisely the $B_s$ flight distance. In this regard, a state-of-the-art pixelized vertex detector is unavoidable.
\item Excellent overall Particle Identification to identify $e^\pm$, $\mu^\pm$ and K$^\pm$. For the latter particles, a Particle Identification system over a large range (at least up to $\sim 25$ GeV) is necessary. One may consider a specific PID system. However the wide momentum range makes this endeavour somewhat difficult. In particular, one has to ensure a minimal amount of material in front of the calorimeter to avoid degrading the energy resolution. An alternative is to use $dE/dx$, using cluster counting combined with an accurate Time of Flight system with a resolution of the order of $10$ ps. A strong R$\&$D program is thus necessary to assess experimentally what can be achieved in this area.

\end{itemize}

\subsubsection{Disentangling the polarization states to maximize $\eta_{\phi\phi}$}
As mentioned in this paper, averaging over the various polarization leads to an effective $\eta_{\phi\phi}$, which is $+0.416$ for $\phi\phi$. In order to improve further the experimental error, one needs to carry out a full angular analysis. This is also very interesting for the determination of
\begin{itemize}
\item The polarization fractions $f_L,\ f_\parallel$ and $f_\perp$ and confront them to QCD Factorization. As mentioned above, one can study CP violation ignoring the angular dependence. In that case, the measure CP violating angle $\phi_{\phi\phi}$ is reduced by the factor $\eta_{\phi\phi} =1-2f_\perp =0.416$. However, to be fully correct, this factor has to be corrected for the detector acceptance as it varies slightly according to the polarization states. 
\item The polarization dependent CP violating phases. Indeed these phases may be different for the various polarization and thus their determination would test with deeper detail the CP sector and eventually be sensitive to BSM physics.

\end{itemize}

\subsection{Implication for New Physics in mixing}	

In the case of FCCee, one has instead of (\ref{6-11bise}),
\beq
{{\rm Im}\Delta_s \over {\rm Re}\Delta_s} \simeq 2 \lambda^2 \eta\ \ {\rm Re}\left({a_c-a_u \over a_c}\right) +\phi_{\phi \phi}^{FCCee} = -\phi_{\phi \phi}^{SM} + \phi_{\phi \phi}^{FCCee}
\label{6-11tere}
\eeq

\noi If FCCee measures the CP phase of the SM within QCDF, i.e. the value (\ref{5-9-2e}),  
\noi $\phi_{\phi \phi}^{SM} \simeq - 2 \lambda^2 \eta\ {\rm Re} \left( {a_c-a_u \over a_c} \right) = -0.011$, then consistently one has ${{\rm Im}\Delta_s} = 0$.\par

If we assume the central value found by LHCb (\ref{5-2e},\ref{5-2-1e}) and the sensitivity $\delta(\phi_{\phi \phi}) = 0.01$, instead of the domain (\ref{6-17bise}) in the case of LHCb,  one would get from (\ref{6-11tere}) the domain for NP in FCCee,
\beq
{{\rm Im}\Delta_s \over {\rm Re}\Delta_s} = 0.011 - 0.073 \pm 0.010 = - 0.062 \pm 0.010
\label{6-3-1e}
\eeq

\noi and to summarize one gets, with the assumed $1\sigma$ error,
\beq
-0.072 \leq {{\rm Im}\Delta_s \over {\rm Re}\Delta_s}  \leq - 0.052
\label{6-3-1e}
\eeq

In Fig. 8 this domain is compared to the one allowed by LHCb. This example shows that the sensitivity at FCCee would be efficient to put NP in evidence.

\qquad \includegraphics[scale=0.9]{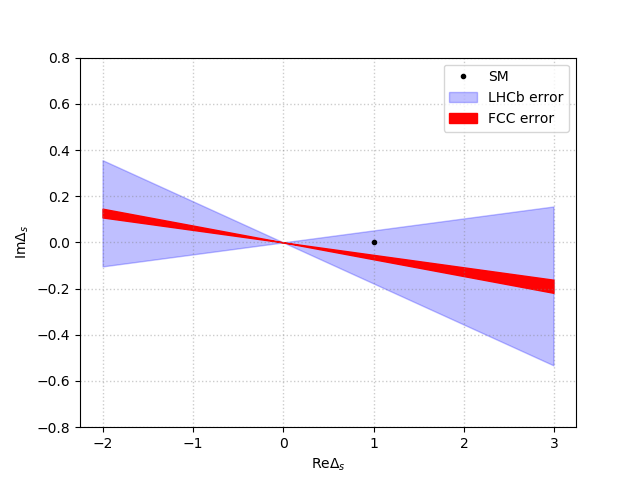}

Fig. 8. The blue area shows the domain allowed in the plane Re$\Delta_s$-Im$\Delta_s$ from the $B_s \to \phi \phi$ data at LHCb and the red area the expectation at FCCee with the central value of LHCb and the assumed uncertainty of Fig. 6 at $1\sigma$. Taking into account the constraints from other observables \cite{CKMfitter}, then only the domain for $Re \Delta_s > 0$ remains.

\vskip 3 truemm

\section{Coments on polarization-dependent rates and CP violation in QCDF}

In this Section we summarize the main aspects of the calculation of the polarization fractions for the Penguin decay $\overline{B}_s \to \phi \phi$ in the QCDF scheme. \par
As argued in Section 2, owing to the hierarchies (\ref{2-2e}), and neglecting the amplitude $A(\overline{B}_s \to \phi \phi, h = +)$ as in (\ref{3e}), one expects the ordering (\ref{4-1bise}), at odds with Table 1. We will see now how QCDF gives a much smaller longitudinal fraction.\par
Let us consider the ratio of transverse to longitudinal amplitudes,
\beq
\label{C-2e}
R = {A(\overline{B}_s \to \phi \phi, h = -) \over A(\overline{B}_s \to \phi \phi, h = 0)}
\eeq

\noi that gives, in the limit  (\ref{3e}), the polarization fractions
\beq
\label{C-3e}
f_L \simeq {1 \over 1\  + \mid R \mid^2} \ , \qquad \qquad f_\parallel \simeq f_\perp \simeq {\mid R \mid^2 \over 2(1\ + \mid R \mid^2)}
\eeq

\noi From the notation (\ref{17bise}), we get
\beq
\label{C-4e}
R = {\sum_{p=u,c}\lambda'_p\ S^{p,-}\ A^-_{\phi \phi} +\ (\lambda'_u + \lambda'_c)\ T^-\ B_{\phi \phi}^- \over \sum_{p=u,c}\lambda'_p\ S^{p,0}\ A^0_{\phi \phi} +\ (\lambda'_u + \lambda'_c)\ T^0\ B_{\phi \phi}^0}
\eeq

\noi where the CKM factors $\lambda'_p$ are given by (\ref{17e},\ref{18e}).\par 
The coefficients for $h = 0$ of the direct diagram $A_{\phi \phi}^0$ and the annihilation diagram $B_{\phi \phi}^0$ are given by (\ref{18-1e},\ref{18-2e}),
\beq
\label{C-4-1}
A^0_{\phi \phi} = i {G_F \over \sqrt{2}}\ m_{B_s}^2 A_0^{B_s \to\phi} (m_{\phi}^2) f_{\phi} \ , \qquad \qquad B^0_{\phi \phi} = i {G_F \over \sqrt{2}} f_{B_s} f_{\phi} f_{\phi}
\eeq

\noi while the coefficients for $h = -$ of the direct diagram $A_{\phi \phi}^-$ and the annihilation diagram $B_{\phi \phi}^-$ are given by
\beq
\label{C-4-2}
A^-_{\phi \phi} = i {G_F \over \sqrt{2}}\ m_{B_s} m_\phi F_-^{B_s \to\phi} (m_{\phi}^2) f_{\phi} \ , \qquad \qquad B^-_{\phi \phi} = i {G_F \over \sqrt{2}} f_{B_s} f_{\phi} f_{\phi}
\eeq

The relevant combinations of the coefficients in QCDF for the $h = 0$ polarization are $S^{p,0} (p = u,c)$ and $T^0$ given by (\ref{22-1e}), $T^0$ being the annihilation.\par
In the paper we have chosen the $h = 0$ polarization and the corresponding CP violation, as made explicit in Section 4.2.\par
We need also the combinations of coefficients for the $h = -$ polarization, namely
$$S^{u,-} = 2\left(a_4^{u,-} + a_3^- +a_5^- - {1 \over 2} (a_7^{u,-} + a_9^{u,-} + a_{10}^{u,-})\right) $$
$$S^{c,-} = 2\left(a_4^{c,-} + a_3^-+a_5^- - {1 \over 2} (a_7^{c,-} + a_9^{c,-} + a_{10}^{c,-})\right) $$
\beq
\label{C-5e}
T^- = 2\left(b_3^- + b_4^- - {1 \over 2}  \left(b_3^{EW,-} + b_4^{EW,-} \right)\right)
\eeq

For the calculation of the $h = -$ combinations (\ref{C-5e}) we use the very explicit formulas of Appendix A of ref. \cite{BENEKE-2007}, that we compute for the $\overline{B}_s \to \phi \phi$ amplitude in our Appendix B.\par
Since the annihilation, as we will see, is crucial for the description of the polarization fractions, we use the model (2) of Table 3 for the $h = 0$ annihilation coefficients, computed in the Appendix C below, that includes the annihilation parameter 
\beq
\label{C-5bise}
X_A = \left(1 + \rho_A e^{i \phi_A} \right)\ \ln {m_{B_s} \over \Lambda_h}
\eeq

\noi consistently with the calculation of the annihilation amplitude for the transverse $h = -$ polarization.\par
For the sake of simplicity, and just to see for the moment the trend of the different effects, we use the central values of the parameters. In particular, we take for the annihilation parameters (\ref{C-5bise}) the following values \cite{BENEKE-2007},
\beq
\label{C-5tere}
\rho_A = 0.6 \ , \qquad \qquad \phi_A = -40^0  \ , \qquad \qquad \Lambda_h = 0.5\ {\rm GeV}
\eeq

\noi Below   we will take into account the possible errors.

\vskip 5 truemm

\subsection{Remark on the form factors}

\vskip 2 truemm

Let us first make a remark on the form factors $A_0^{B_s \to\phi}$ and $F_-^{B_s \to\phi}$. In \cite{BENEKE-2007} the form factors used are consistent with the values from the QCD Sum Rules calculation of \cite{BALL-BRAUN}, that are somewhat different from the ones computed later in \cite{BALL-ZWICKY-2}, and used in ref. \cite{BUCHALLA} for the longitudinal amplitude. Both choices lead to different coefficients $A^0_{\phi \phi}$ and $A^-_{\phi \phi}$ in (\ref{C-4-1},\ref{C-4-2}). \par
To see the trend of the effect of the form factors on the polarization fractions, we will perform the calculation of the ratio (\ref{C-4e}) for the two choices of the form factors used in \cite{BENEKE-2007} and \cite{BUCHALLA}, using for the moment the central values
\beq
\label{C-5-1e}
(1) \qquad A_0^{B_s \to \phi}(0) = 0.38 \ , \qquad \qquad F_-^{B_s \to \phi}(0) = 0.65 \qquad
\eeq
\beq
\label{C-5-2e}
(2) \qquad A_0^{B_s \to \phi}(0) = 0.47\ , \qquad \qquad F_-^{B_s \to \phi}(0) = 0.72 \qquad
\eeq

\noi The last number is obtained from \cite{BENEKE-2007}
\beq
\label{C-6bise}
F_-^{B_s \to\phi} (0) = \left(1 + {m_\phi \over m_{B_s}} \right) A_1^{B_s \to\phi} (0) + \left(1 - {m_\phi \over m_{B_s}} \right) V^{B_s \to\phi} (0) = 0.72
\eeq

\noi where, to be consistent, we have used the values for $A_1^{B_s \to\phi} (0) = 0.31$ and $V^{B_s \to\phi} (0) = 0.43$ from the same sum rules calculation \cite{BALL-ZWICKY-2} giving $A^0_{\phi \phi} (0) = 0.47$, used in \cite{BUCHALLA} and Table 2 for the longitudinal amplitude.\par
Let us remark that there is a sizeable difference for $A_0^{B_s \to \phi}(0)$ between both estimations (\ref{C-5-1e}) and (\ref{C-5-2e}).\par
For the two choices (1) and (2), (\ref{C-5-1e},\ref{C-5-2e}), we get the cofficients $A^0_{\phi \phi}$ and $A^-_{\phi \phi}$ (\ref{C-4-1},\ref{C-4-2}),
\beq
\label{C-6-2e}
(1) \qquad A^0_{\phi \phi} = i {G_F \over \sqrt{2}} \times (2.422\ {\rm GeV^3}) \ , \qquad A^-_{\phi \phi} = i {G_F \over \sqrt{2}} \times (0.787\ {\rm GeV^3})
\eeq
\beq
\label{C-6-3e}
(2) \qquad A^0_{\phi \phi} =  i {G_F \over \sqrt{2}} \times (2.995\ {\rm GeV^3})  \ , \qquad  A^-_{\phi \phi} =  i {G_F \over \sqrt{2}} \times (0.872\ {\rm GeV^3})  
\eeq

\vskip 4 truemm

Let us split the complete calculation of the ratio (\ref{C-4e}) into different steps with different physical meaning in order to see their particular effect on the final result.\par

\vskip 8 truemm

\subsection{Limit of Naive Factorization}

\vskip 2 truemm

We observe first that the ratios of the coefficients of direct diagrams for $h = -$ with respect to $h = 0$ are given by
\beq
\label{C-7-1e}
\left( {A^-_{\phi \phi} \over A^0_{\phi \phi}} \right)^{(1)} = 0.325 \ , \qquad \qquad \left( {A^-_{\phi \phi} \over A^0_{\phi \phi}} \right)^{(2)} = 0.291
\eeq

\noi respectively for the choices (1) and (2).\par

Therefore, at this step one would naively expect a much larger longitudinal polarization fraction than a transverse one.\par 
In NF we take the limits $a_i^{c,h} = a_i^{u,h} \to C_i + {C_{i\pm1} \over N_c}$ (upper sign for $i$ odd and lower sign for $i$ even) and $b_i^h \to 0$. Then we find a longitudinal fraction 
\beq
\label{C-7-1e}
f_L^{NF} = {\mid A^0_{\phi \phi}\mid^2 \over \mid A^0_{\phi \phi}\mid^2 + \mid A^-_{\phi \phi}\mid^2}
\eeq

\noi that gives
\beq
\label{C-7-2bise}
\left( f_L^{NF} \right)^{(1)} \simeq 0.90 \ , \qquad \qquad \left( f_L^{NF} \right)^{(2)} \simeq 0.92
\eeq

\noi respectively for the choices (1) and (2).

\vskip 10 truemm

\subsection {Coefficients in QCDF for helicities $h = 0$ and $h = -$}

In Tables 2 and 3 we have given the coefficients for the $h = 0$ amplitude. For the annihilation coefficients for $h = 0$ we will consider here Model (2), consistently with the annihilation in the $h = -$ amplitude that we give here below.\par

\vskip 5 truemm

 In Tables 7 and 8 we give the results of the QCDF coefficients for the $h = -$ amplitude, computed in Appendix B.\par 

\vskip 8 truemm

\begin{center}
\begin{tabular}{|c|c|c|}
\hline 
$a_3^-+a_5^-$ & $a_4^{u,-}$ & $a_4^{c,-}$ \\ \hline
$-0.005-0.001 i$ & $-0.047-0.015 i$ & $-0.046-0.001 i$ \\ \hline
\end{tabular}

\begin{tabular}{|c|c|c|c|}
\hline 
$(a_7^{u,-}+a_9^{u,-})/ \alpha$ & $(a_7^{c,-}+a_9^{c,-})/ \alpha$ & $a_{10}^{u,-}/ \alpha$ & $a_{10}^{c,-}/ \alpha$ \\ \hline
$0.68-0.03 i$ & $0.62-0.03 i$ & $0.29+0.17 i$ & $0.29+0.18 i$ \\ \hline
\end{tabular}

\vskip 4 truemm

Table 7. Coefficients at NLO obtained in QCD Factorization for helicity $h = -$.

\end{center}

\begin{center}
\begin{tabular}{|c|c|c|c|}
\hline 
$r_A b_3^-$ & $r_A b_4^-$ & $r_A b_3^{EW,-}/\alpha$ & $r_A b_4^{EW,-}/\alpha$ \\ \hline
-0.018+0.018 i & -0.001 & 0.022-0.028 i & 0.002 \\ \hline
\end{tabular}
\end{center}

\noi Table 8. Coefficients for $h = -$ annihilation, with $r_A = B_{\rho \rho}/A_{\rho \rho} = 5. \times 10^{-3}$, from $f_{B_d} = 0.200$ GeV, $f_\rho = 0.209$ GeV, $A_0^{B_d \to \rho}(0) = 0.30$ and $\alpha = 1/129$ \cite{BUCHALLA}.\par

\vskip 3 truemm

We will use the coefficients of Tables 2 and 3 for $h = 0$ and Tables 7 and 8 for $h = -$ to compute the ratio $A(\overline{B}_s \to \phi \phi, h = -) / A(\overline{B}_s \to \phi \phi, h = 0)$ and therefore $f_L$. In order to clarify the origin of the enormous shift between the large naive value (\ref{C-7-2bise}) and the final QCDF results and the data, we consider now different steps of the calculation.\par
Notice that we use universal values for both sets of form factors (\ref{C-5-1e}, \ref{C-5-2e}). One must notice however, that the Hard scattering contributions depend on the form factors. We will however neglect small corrections coming from the difference of the form factors in these contributions.

\vskip 4 truemm

\subsection {Limit of neglecting Annihilation in QCD Factorization}

\vskip 3 truemm

To further clarify the origin of the final result for $f_L$ we take the formal limit ${\rm Annihilation} \to 0$ in the QCDF expression (\ref{C-4e}), and we find
$$\left({\mid A(\overline{B}_s \to \phi \phi, h = -) \mid \over \mid A(\overline{B}_s \to \phi \phi, h = 0)\mid} \right)^{(1)}_{Ann \to 0} = 0.606$$
\beq
\label{C-9e}
\left({\mid A(\overline{B}_s \to \phi \phi, h = -) \mid \over \mid A(\overline{B}_s \to \phi \phi, h = 0)\mid} \right)^{(2)}_{Ann \to 0} = 0.526
\eeq

\vskip 3 truemm 

\noi for the choices of form factors (1) and (2), (\ref{C-5-1e},\ref{C-5-2e}).\par
We realize that $f_L$ decreases with respect to Naive Factorization, but still remains large,
\beq
\label{C-9-2e}
\left(f_L\right)^{(1)}_{Ann \to 0} = 0.73 \ , \qquad \qquad \left(f_L\right)^{(2)}_{Ann \to 0} = 0.78
\eeq

\noi for both choices of the form factors.\par

Moreover, it is important to emphasize that the parametrization (\ref{C-5bise}) of the divergence in the annihilation plays an important role in the final result. Indeed, it is illustrative to compute the theoretical expression for $f_L$ in the formal limit $\rho_A = 0$, i.e. for the default value $X_A = \ln{m_{B_s} \over \Lambda_h}$
\beq
\label{C-9-3e}
\left(f_L\right)^{(1)}_{\rho_A \to 0} = 0.81 \ , \qquad \qquad \left(f_L\right)^{(2)}_{\rho_A \to 0} = 0.68
\eeq

\noi These values are much larger than experiment (Table 1), due to the cancellation $b_3^0+b_4^0 \simeq 0$ (Table 3, Model (1)). One can also see in (\ref{C-9-3e}) the effect of the two choices of the form factors.

\vskip 5 truemm

\subsection {Final results in QCD Factorization}

\vskip 3 truemm

In the QCDF scheme, it is the Annihilation contribution that mostly changes the pattern of the polarization fractions. What happens is that the Annihilation for $h = -$ interferes constructively with the rest of the contributions. On the contrary, for $h = 0$ the annihilation amplitude has the opposite sign, and interferes destructively.\par

Taking all contributions into account, including Annihilation, our final result for the ratio of the moduli of $h = -$ and $h = 0$ amplitudes is
\beq
\label{C-11e}
\left( {\mid A(\overline{B}_s \to \phi \phi, h = -) \mid \over \mid A(\overline{B}_s \to \phi \phi, h = 0)\mid} \right)^{(1)} = 1.253 \ , \qquad 
\left( {\mid A(\overline{B}_s \to \phi \phi, h = -) \mid \over \mid A(\overline{B}_s \to \phi \phi, h = 0)\mid} \right)^{(2)} = 1.086
\eeq

\noi giving the polarization fractions
$$f_L^{(1)} = 0.389 \ , \qquad \qquad f_\parallel^{(1)} = f_\perp^{(1)} = 0.305$$
\beq
\label{C-12e}
f_L^{(2)} = 0.456 \ , \qquad \qquad f_\parallel^{(2)} = f_\perp^{(2)} = 0.272
\eeq

\noi for both choices of the form factors. The results for the choice (1) is in qualitative agreement with the experimental results of Table 1, while for the choice (2) the polarization fractions are somewhat off.\par 
The $BR$ of the decays into definite helicity vector mesons read,
\beq
\label{19bise}
BR(\overline{B} \to V_1 V_2, h) = S {\tau_B \over {8 \pi m_B^2}} \mid A(\overline{B} \to V_1 V_2, h) \mid^2 p
\eeq

\noi where $S = 1/2$\ for identical particles, like in the case $B_s \to \phi \phi$, or $S = 1$ otherwise.

\noi Adding the rates for the $h = 0$ and $h = -$ polarizations, we find the total branchig ratio, 
\beq
\label{C-13e}
BR(\overline{B}_s \to \phi \phi)^{(1)} = 24.8 \times 10^{-6} \ , \qquad \qquad BR(\overline{B}_s \to \phi \phi)^{(2)} = 29.7 \times 10^{-6}
\eeq

\noi for both choices of the form factors. Choice (1) is in qualitative agreement with the experimental value of Table 1, and choice (2) gives a value that is too large. We will take into account errors en these quantities below.\par

In summary, the results (\ref{C-11e}-\ref{C-13e}) follow from the constructive interference of the Annihilation in the $h = -$ amplitude, and the destructive interference in the $h = 0$ one.\par

For completeness, we give here the CP violation parameters that we find for the different polarizations, using the notation (\ref{5-1e}) and the choice (1) of the form factors,
$$\mid \lambda_{\phi \phi}^{(L)} \mid\ = 1.015 \ , \qquad \qquad \phi_{\phi \phi}^{(L)} \simeq 10^{-3}$$
\beq
\label{C-14e}
\mid \lambda^{(\parallel)}_{\phi \phi} \mid\ =\ \mid \lambda^{(\perp)}_{\phi \phi} \mid\ = 1.004 \ , \qquad \qquad \phi^{(\parallel)}_{\phi \phi} = \phi^{(\perp)}_{\phi \phi} \simeq 10^{-3} 
\eeq

\noi where the equalities between $\parallel$ and $\perp$ quantities follow from the motivated hypothesis (\ref{3e}).\par

\subsection {Including errors in the calculation of the observables}

For the sake of simplicity we have used the central values of the different parameters, giving the numerical results (\ref{C-12e},\ref{C-13e},\ref{C-14e}) for the observables.\par 
We now take into account uncertainties on these parameters, given in Table 9. It is difficult to estimate the errors on the QCDF coefficients $a_i^h$, and we make the simple guess of a $\pm 10 \%$ error. Concerning the annihilation, in both $h = 0, -$ amplitudes we adopt the errors on the parameters $\rho_A, \phi_A$ of ref. \cite{BENEKE-2007}.\par 
We restrict ourselves to the model (1) for the form factors, eq. (\ref{C-5-1e}).\par
We consider flat distributions around the central values of the parameters, with correlated guessed errors on the $a_i^h (h = 0, -)$. The errors are then added in quadrature and one gets the domains for the different observables of Figs. 9 - 12.

\begin{center}
\begin{tabular}{|c|c|}
\hline 
Parameter & Value \\ \hline
$f_{B_s}$ & $0.230 \pm 0.005\ {\rm GeV}$ \\ \hline
$f_\phi$ & $0.221 \pm 0.005\ {\rm GeV}$ \\ \hline
$A_0^{B_s \to \phi}(0)$ & $0.38 \pm 0.05$ \\ \hline
$F_-^{B_s \to \phi}(0)$ & $0.65 \pm 0.06$ \\ \hline
$a_i^0$ & $(1 \pm 0.10) \times$Values of Table 2, Model (2) \\ \hline
$a_i^-$ & $(1 \pm 0.10) \times$Values of Table 7 \\ \hline
$\rho_A$ & $0.6 \pm 0.2$ \\ \hline
$\phi_A$ & $(- 40 \pm 10) ^0$\\ \hline
\end{tabular}

\end{center}

\noi Table 9. Values and errors of the parameters. For the annihilation in both amplitudes $h = 0,\ h = -$, only large errors for the important parameters $\rho_A$ and $\phi_A$ are considered. 

\includegraphics[scale=0.9]{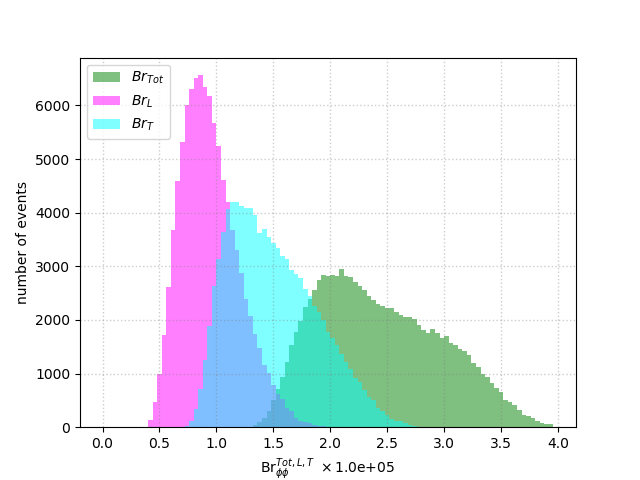}

\noi Fig. 9. L, T and total Branching Ratios with the input parameters of Table 9. 

\includegraphics[scale=0.9]{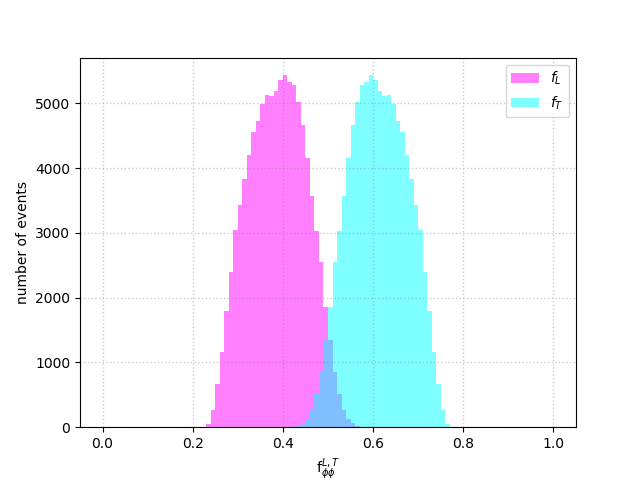}

\noi Fig. 10. L and T polarization fractions with the input parameters of Table 9. 

\noi \includegraphics[scale=0.7]{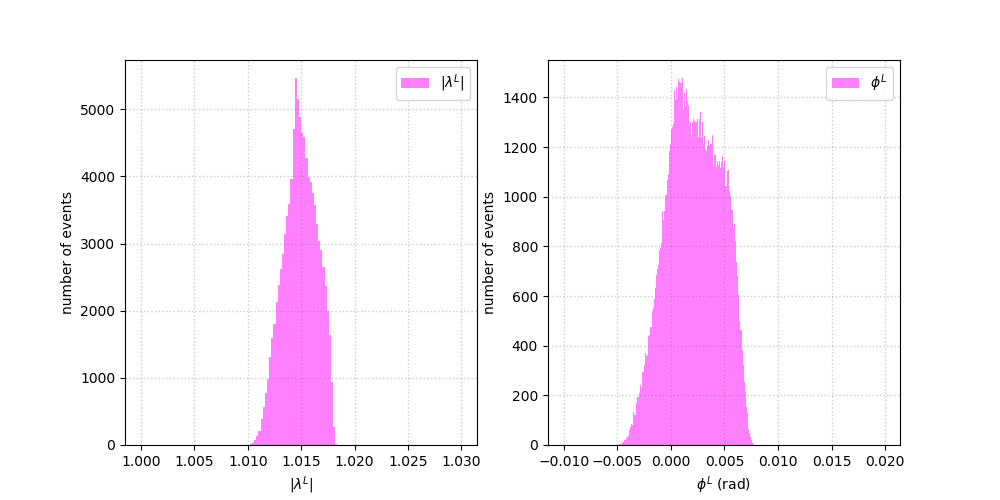} \qquad \qquad \par 
Fig. 11. Modulus and phase of the CP parameter $\lambda^L = {q \over p} {A(\overline{B}_s \to \phi \phi, L) \over A(B_s \to \phi \phi, L)}$ with the input parameters of Table 9. 

\noi \includegraphics[scale=0.7]{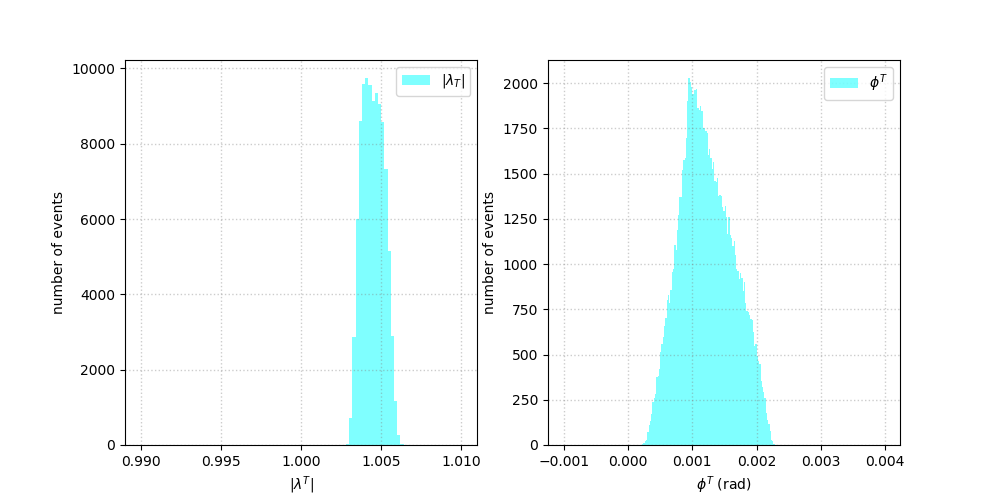} \qquad \qquad \par
\noi Fig. 12. Modulus and phase of the CP parameter $\lambda^T = {q \over p} {A(\overline{B}_s \to \phi \phi, T) \over A(B_s \to \phi \phi, T)}$ with the input parameters of Table 9. 

\vskip 3 truemm

The final result for the fitted observables is given in Table 10.

\vskip 5 truemm

\begin{center}
\begin{tabular}{|c|c|c|}
\hline 
Observable & Result of the fit & Data \\ \hline
$BR$ & $(2.44 \pm 0.63) \times 10^{-5}$ & $(1.85 \pm 0.17) \times 10^{-5}$ \\ \hline
$BR_L$ & $(0.94 \pm 0.24) \times 10^{-5}$ & $(0.70 \pm 0.02) \times 10^{-5}$ \\ \hline
$BR_T$ & $(1.50 \pm 0.41) \times 10^{-5}$ & $(1.15 \pm 0.02) \times 10^{-5}$\\ \hline
$f_L$ & $0.39 \pm 0.08$ & $0.38 \pm 0.01$ \\ \hline
$f_T$ & $0.61 \pm 0.08$ & $0.62 \pm 0.01$ \\ \hline
$\mid \lambda^L \mid$ & $1.015 \pm 0.002$ & $0.99 \pm 0.05$ \\ \hline
$\phi^L$ & $0.002 \pm 0.003$ & $-0.073 \pm 0.115$\\ \hline
$\mid \lambda^T \mid$ & $1.005 \pm 0.001$ & $0.99 \pm 0.05$ \\ \hline
$\phi^T$ & $0.001 \pm 4. \times 10^{-4}$ & $-0.073 \pm 0.115$ \\ \hline
\end{tabular}

\end{center}

\vskip 2 truemm

\noi Table 10. Results of the fit for the observables of $\overline{B}_s \to \phi \phi$. The experimental CP violation parameters correspond to the polarization-independent fit of LHCb \cite{LHCbPHIPHI}.

\vskip 5 truemm

We observe that the branching ratios and polarization fractions are in agreement with the data within $1\sigma$, and the CP violation parameters are close to the Naive Factorization expectations.\par 
On the other hand, since we have worked within the hypothesis that the amplitude $A_+$ vanishes (\ref{3e}), the transverse parallel and transverse perpendicular polarization fractions of the model are equal and given by
\beq
\label{C-14bise}
f_\parallel = f_\perp = 0.305 \pm 0.005
\eeq

\noi consistently with the experimental value of Table 1,
\beq
\label{C-14tere}
f_\perp = 0.29 \pm 0.01
\eeq

\subsection{The decay mode $\overline{B}_d \to \phi \phi$}

Finally, let us remark that the decay mode $\overline{B}_d \to \phi \phi$, for which there is only an experimental upper bound, is very interesting since it follows uniquely from penguin annihilation \cite{BUCHALLA} and could thus be a subtle test of the QCDF scheme. For this mode we find that the longitudinal amplitude dominates very much ($f_L \sim 99 \% $) and we find a branching ratio $\sim (1.2 \pm 0.3) \times 10^{-8}$ consistent with the experimental limit $< 2.7 \times 10^{-8}$.\par

\vskip 5 truemm

\section{Caution remarks and conclusion}

We have assumed NP contributions only for the mixing, as it is usually done. One should discuss however how possible NP in the Penguin diagrams could alter our results. There are a number of arguments pointing to dominance of a possible NP contribution in mixing. For example, assuming a new heavy $W'$ boson, its contribution to the mixing will be quadratic in its mass, while it will be only logarithmic for the Penguin loop. But this deserves to be examined in more detail. The NP contributions in the Penguin decay amplitude are presumably small, but one can already say that, due to these contributions, the complex parameter $\Delta_s$ cannot be exactly the same for the mode $\overline{B}_s \to \phi \phi$ and for the standard mode $\overline{B}_s \to J/\psi \phi$.\par 
It is important also to remark that, although both modes involve the same CP phase $- 2\beta_s$ generated from the mixing, this phase is cancelled by $2\beta_s$ from the decay in the mode $\phi \phi$, while it remains $-2 \beta_s$ in the interference between mixing and decay in the mode $J/\psi \phi$.\par 

On the theoretical side, we have restricted ourselves to the simpler case considered at LHCb \cite{LHCbPHIPHI}, namely the time-dependent mixing induced CP violation, assuming that it is {\it the same} for the different polarization amplitudes, namely longitudinal, transverse parallel and transverse perpendicular. To estimate $\overline{A}/A$ in the SM we have chosen the longitudinal polarization, on which very explicit results are given by \cite{BUCHALLA} within the QCDF scheme, going thus beyond NF.\par
It would be necessary to have $\lambda_{\phi \phi}$ for the different polarizations $L,\parallel, \perp$ in the QCDF scheme. We have performed this calculation along the lines of \cite{BENEKE-2007}, obtaining the results of Table 10, approximately confirming that the LHCb polarization independent fit makes sense.\par
On the other hand, we have peformed a calculation of the polarization fractions in the QCDF scheme for the penguin mode $\overline{B}_s \to \phi \phi$, and we have obtained theoretical values that turn out to be in agreement with experiment. We have traced the contradiction between the large Naive Factorization value for the longitudinal fraction $f_L$, close to $f_L \sim 0.90$, and the final much lower theoretical value within QCDF, closer to experiment, $f_L \sim 0.40$. We have paid special attention to the different contributions to $f_L$, and we have arrived to the conclusion that the annihilation for both the $h = 0$ and $h = -$ helicities is crucial in reproducing the data. We point out that the parametrization (\ref{C-5bise}) of the weak annihilation IR-divergent quantity plays a crucial role in the final numerical result. One must keep in mind that this is a model-dependent quantity, and therefore the theoretical description of the polarization fractions largely relies on a model.\par

In conclusion, we have analyzed the pure Penguin mode to two vector mesons $\overline{B}_s \to \phi \phi$, that is experimentally clean and for which there are already measurements at LHCb. We have emphasized that in the interference quantity $\lambda_{\phi \phi} = \left({q \over p}\right)_{B_s} {A(\overline{B}_s \to \phi \phi) \over A(B_s \to \phi \phi)}$, in the Naive Factorization limit the CP phase from the mixing exactly cancels the CP phase from the decay ratio. Therefore, this mode is suitable to distinguish possible NP in mixing. We estimate the deviation from the cancellation between the CP phases in mixing and decay by analyzing the ratio ${A(\overline{B}_s \to \phi \phi) \over A(B_s \to \phi \phi)}$ beyond Naive Factorization, making use of the QCD Factorization scheme, that applies to this case of two light mesons in the final state. We compare the theoretical value obtained for the modulus and phase of $\lambda_{\phi \phi}$ to the measurement at LHCb, and its implications for NP.\par
Finally, we make an estimation of the expected sensitivity at the future FCCee experiment for the CP phase and modulus of $\lambda_{\phi \phi}$. We find $\delta(\mid \lambda_{\phi \phi}\mid) = 0.004$ and $\delta(\phi_{\phi \phi}) = 0.009$ rad and, comparing to the LHCb data, we point out the implications at FCCee in the search of NP.

\vskip 10 truemm

\noi {\Large \bf Appendix A. Wilson operators and short distance coefficients}

\vskip 3 truemm

The effective Lagrangian is given by
\beq
H = {G_F \over \sqrt{2}} \left[V^*_{cs} V_{cb} (C_1 O_1 + C_2 O_2) - V^*_{ts} V_{tb} (C_3 O_3 + C_4 O_4  + C_5 O_5 + C_6 O_6) \right] + H_{EW}^{Penguin}
\label{A-1e}
\eeq
$$O_1 = \left[\overline{c} \gamma^\mu (1-\gamma_5) b \right] \left[\overline{s} \gamma_\mu  (1-\gamma_5) c \right]$$
$$O_2 = \left[\overline{c}_\alpha \gamma^\mu (1-\gamma_5) b_\beta \right] \left[\overline{s}_\beta \gamma_\mu  (1-\gamma_5) c_\alpha \right]$$
$$O_3 = \left[\overline{s} \gamma^\mu (1-\gamma_5) b \right] \left[\overline{q}\gamma_\mu  (1-\gamma_5) q \right]$$
\beq
O_4 = \left[\overline{s}_\alpha \gamma^\mu (1-\gamma_5) b_\beta \right] \left[\overline{q}_\beta \gamma_\mu  (1-\gamma_5) q_\alpha \right]
\label{A-2e}
\eeq
$$O_5 = \left[\overline{s} \gamma^\mu (1-\gamma_5) b \right] \left[\overline{q}\gamma_\mu  (1+\gamma_5) q \right]$$
$$O_6 = \left[\overline{s}_\alpha \gamma^\mu (1-\gamma_5) b_\beta \right] \left[\overline{q}_\beta \gamma_\mu  (1+\gamma_5) q_\alpha \right]$$

\noi where the Penguin contribution reads
\beq
H_{EW}^{Penguin} = - {G_F \over \sqrt{2}} V^*_{ts} V_{tb} (c_7 O_7 + c_8 O_8  + c_9 O_9 + c_{10} O_{10})
\label{A-3e}
\eeq
$$O_7 = {3 \over 2} \left[\overline{s} \gamma^\mu (1-\gamma_5) b \right] \left[e_q \overline{q}\gamma_\mu  (1+\gamma_5) q \right]$$
\beq
O_8 = {3 \over 2} \left[\overline{s}_\alpha \gamma^\mu (1-\gamma_5) b_\beta \right] \left[e_q \overline{q}_\beta \gamma_\mu  (1+\gamma_5) q_\alpha \right]
\label{A-4e}
\eeq
$$O_9 = {3 \over 2} \left[\overline{s} \gamma^\mu (1-\gamma_5) b \right] \left[e_q \overline{q}\gamma_\mu  (1-\gamma_5) q \right]$$
$$O_{10} = {3 \over 2} \left[\overline{s}_\alpha \gamma^\mu (1-\gamma_5) b_\beta \right] \left[e_q \overline{q}_\beta \gamma_\mu  (1-\gamma_5) q_\alpha \right]$$

\noi and the Wilson coefficients are given in Table 11.

\begin{center}
\begin{tabular}{|c|c|c|c|c|}
\hline 
$C_1$ & $C_2$ & $C_3$& $C_4$ & $C_5$\\ \hline
$1.086$ & $-0.191$ & $0.014$ & $-0.035$ & $0.010$ \\ \hline \hline
$C_6$ & $C_7$ & $C_8$ & $C_9$ & $C_{10}$ \\ \hline
$-0.043$ & $0.011\ \alpha$ & $0.059\ \alpha$ & $-1.229\ \alpha$ & $0.246\ \alpha$ \\ \hline
\end{tabular}
\end{center}
Table 11. Wilson coefficients at NLO for $\mu = m_b = 4.2\ {\rm GeV}$ in the NDR scheme.

\vskip 10 truemm

\noi {\Large \bf Appendix B. The amplitude $A(\overline{B}_s \to \phi \phi, h = -)$}

\vskip 5 truemm
Here we make explicit some details of the calculation of the $h = -$ QCDF coefficients $a_i^-$ and $b_i^-$ summarized in Tables 9 and 10. 

\vskip 10 truemm

\noi {\Large \bf The coefficients $a_i^-$}

\vskip 3 truemm

From formulas (11) of \cite{KAGAN} and (53) of \cite{BENEKE-2007} the coefficients $a_i^{p,-} (p = u, c)$ write

$$a_i^{p,-} = \left(C_i(\mu) + {C_{i\pm 1}(\mu) \over N_c} \right) N_i^- + {C_{i\pm 1}(\mu) \over N_c} {C_F \alpha_s(\mu) \over {4 \pi}} V_i^-$$
\beq
\label{B-1e} 
+ {C_{i\pm 1}(\mu_h) \over N_c} {C_F \alpha_s(\mu_h) \over {4 \pi}} {4 \pi^2 \over N_c} H_i^- + P_i^{p,-}
\eeq

\noi where we have made explicit that the scale is not the same for the Hard scattering term. The coupling $\alpha_s$ and the Wilson coefficients for both scales are given respectively by (\ref{C-7-3e})  and Tables 11 and 12.

\vskip 10 truemm

\noi {\bf \large Two scales $\mu$ and $\mu_h$ in QCDF}

\vskip 5 truemm

Let us make an important remark about the scales considered in QCDF.\par 
The Wilson coefficients $C_i(\mu)$ relevant to the Vertex $V$ and Penguin $P$ contributions and the order $\alpha_s(\mu)$ NLO corrections specific to QCDF are estimated at some scale $\mu$, that for definiteness we take to be $\mu = m_b = 4.2\ {\rm GeV}$.\par
For the Hard scattering $H$, due to off-shellness of the gluon in these diagrams \cite{BBNS-01}, one adopts {\it a different scale} $\mu_h = \sqrt{\Lambda_h \mu}$ with $\Lambda_h = 0.5\ {\rm GeV}$ and $\mu = m_b$, evaluating both at this scale the Wilson coefficients $C_i(\mu_h)$ and the coupling $\alpha_s(\mu_h)$ specific to the NLO corrections of QCDF. For the Annihilation corrections $A$ one adopts the same Wilson coefficients $C_i(\mu_h)$ and coupling $\alpha_s(\mu_h)$.\par

This important point, that has sizeable quantitative consequences, has been put forward in refs. \cite{BENEKE-2007} and \cite{BUCHALLA}, following the earlier paper by Beneke and Neubert \cite{BENEKE-NEUBERT}, who underline that both the running coupling constant and the Wilson coefficients for the Hard scattering and Annihilation terms have to be evaluated at an intermediate scale $\mu_h \sim \sqrt{\Lambda_{QCD} m_b}$ rather than $\mu \sim m_b$.\par

This different scale $\mu_h$ is much lower than $\mu$, namely
\beq
\label{C-7-2e}
\mu = m_b = 4.2\ {\rm GeV} \ , \qquad \Lambda_h = 0.5\ {\rm GeV} \ , \qquad \mu_h = \sqrt{\mu \Lambda_h} = 1.45\ {\rm GeV} 
\eeq

\noi This increases substantially $\alpha_s(\mu_h)$ and also the Wilson coefficients evaluated at this scale. One has 
\beq
\label{C-7-3e}
\alpha_s(4.2) = 0.224 \ , \qquad \qquad \alpha_s(1.45) = 0.338
\eeq

\vskip 3 truemm

\noi and the Wilson coefficients at this latter scale are given in Table 12 (for the formalism see the detailed paper \cite{BUCHALLA-HARLANDER-1995}). \par

\vskip 5 truemm

\begin{center}
\begin{tabular}{|c|c|c|c|c|}
\hline 
$C_1$ & $C_2$ & $C_3$& $C_4$ & $C_5$\\ \hline
$1.190$ & $-0.373$ & $0.027$ & $-0.062$ & $0.012$ \\ \hline \hline
$C_6$ & $C_7$ & $C_8$ & $C_9$ & $C_{10}$ \\ \hline
$-0.086$ & $0.003\ \alpha$ & $0.128\ \alpha$ & $-1.353\ \alpha$ & $0.430\ \alpha$ \\ \hline
\end{tabular}
\end{center}

Table 12. Wilson coefficients at NLO in the NDR scheme, at the scale $\mu_h = \sqrt{\mu \Lambda_h} = 1.45\ {\rm GeV}$, relevant for the Hard scattering and Annihilation contributions.

\vskip 3truemm

Clearly, the Hard scattering and Annihilation diagrams will be very much enhanced due to the low scale $\mu_h$.

\vskip 8 truemm

\noi {\bf \large Phenomenological parametrization of divergences}

\vskip 3 truemm

There are three divergences in QCDF, for which one adopts the phenomenological parametrizations of ref. \cite{BENEKE-2007}, with $\Lambda_{QCD} = 0.225\ {\rm GeV}$ and $\Lambda_h = 0.5\ {\rm GeV}$,
\beq
\label{C-9-1e}
X_H = {\rm ln}\left( {m_{B_s} \over \Lambda_h}\right) \ , \qquad X_A = \left(1+0.6\ e ^{-i 40^0} \right) {\rm ln}\left( {m_{B_s} \over \Lambda_h}\right) \ , \qquad X_L = {m_b \over \Lambda_{QCD}} 
\eeq

\noi $X_H$ enters in the Hard scattering, and $X_A$ and $X_L$ in the Annihilation contributions to the $h = -$ amplitude.\par

\vskip 8 truemm

\noi {\large \bf Normalization factors}
\beq
\label{B-2e}
N_i^-(\phi) = 0\ \ (i = 6, 8) \ , \qquad \qquad N_i^-(\phi) = 1\ \ (i \not = 6, 8)
\eeq

\vskip 3 truemm

\noi {\large \bf Vertex corrections}

\vskip 5 truemm

For $\mu = m_b$ one gets the simplified formulas

$$\ \ \ \ V_i^- = \int_0^1 dy \phi_{b}(y) \left[- 18 + g_T(y) \right] \ , \qquad \qquad i \in \{1, 2, 3, 4, 9, 10 \}$$
\beq
\label{B-3e}
\qquad \ \ \ \ \ \ V_i^- = \int_0^1 dy \phi_{a}(y) \left[6 - g_T(y) \right] \ , \qquad \qquad \ \ \ \ i \in \{5, 7 \} \qquad \qquad
\eeq
$$\qquad \ \ \ V_i^- = 0 \ , \ \ \qquad \qquad \qquad \qquad \qquad \qquad \qquad \ \  i \in \{6, 8 \} \ \ \qquad \qquad $$

\vskip 3 truemm

with
$$g_T(y) = {4-6 y \over 1-y} \ln y - 3 i \pi$$ 
\beq
\label{B-4e}
+ \left (2 {\rm Li}_2(y) - \ln^2 y + {2 \ln y \over 1-y} - (3 + 2 \pi i) \ln y - [y \to 1-y ] \right)
\eeq

\vskip 3 truemm

Keeping only the lower twist, the distribution functions read
\beq
\label{B-5e}
\Phi_V(v) = 6v(1-v)
\eeq
\beq
\label{B-6e}
\phi_a(u) = \int^1_u dv {\Phi_V(v) \over v}  = 3(1-u)^2\ , \qquad \phi_b(u) = \int^u_0 dv {\Phi_V(v) \over 1-v} = 3u^2
\eeq

\vskip 3 truemm

Numerically, one gets, for $\mu = m_b$,
$$\ V_i^-(\phi) = -14 - 6i\pi \ , \qquad \qquad i \in \{1, 2, 3, 4, 9, 10 \}$$
\beq
\label{B-7e}
V_i^-(\phi) = 2 + 6i\pi \ , \qquad \qquad \ \ \ \ \ i \in \{5, 7 \} \qquad \qquad
\eeq
$$\ V_i^-(\phi) = 0 \ , \qquad \qquad \qquad \qquad i \in \{6, 8 \} \qquad \qquad $$

\vskip 30 truemm

\noi {\large \bf Penguin corrections} 

\vskip 5 truemm

Formulas (57)-(59) of \cite{BENEKE-2007} simplify for $\mu = \nu = m_b$, 
$$P_6^{-,p} = P_8^{-,p} = 0 \qquad (p = u, c)$$
$$P_4^{-,p} = {\alpha_s C_F \over 4 \pi N_c}\ \{ C_1 \left[{2 \over 3} - G^-(s_p) \right] + C_3 \left[{4 \over 3} - G^-(0) - G^-(1)\right]$$
\beq
\label{B-8e}
+\ (C_4 +C_6) \left[-3 G^-(0) - G^-(s_p) - G^-(1)\right] \}
\eeq
$$P_7^{-,p} = P_9^{-,p} = - {\alpha \over 3 \pi}\ C_{7\gamma}^{eff} {m_{B_s}^2 \over m_\phi^2} + {2 \alpha \over 27 \pi} (C_1 + N_c C_2)\left[\delta_{pc} \ln {m_c^2 \over m_b^2} +1 \right]$$
$$P_{10}^{-,p} = {\alpha \over 9 \pi N_c} (C_1 + N_c C_2) \left[{2 \over 3} - G^-(s_p) \right]$$

\noi The enhancement factor ${m_{B_s}^2 \over m_\phi^2}$ modifies the naive power counting in the EW penguins.\par
The function $G^-(s)$ reads
\beq
\label{B-9e}
G^-(s) = \int^1_0 dy\ \phi_b(y) G(s-i\epsilon,1-y)
\eeq

\noi where $G(s,y)$ is the penguin function, formula (51) of \cite{BBNS-01},
\beq
\label{B-10e}
G(s,x) = {2(12s + 5x - 3x\ln s) \over 9x} - {4 \sqrt{4s-x} (2s+x) \over 3 x^{3/2}} \arctan\sqrt{{x \over 4s-x}}
\eeq

\noi From $m_c = 1.3\ {\rm GeV}, m_b = 4.2\ {\rm GeV}$ one has $s_c = 0.958$ and the values 
\beq
\label{B-11e}
G^-(0) = 2.333+2.094\ i \ , \qquad G^-(1) = 0.035 \ , \qquad G^-(s_c) = 2.090+0.315\ i 
\eeq

\noi From (\ref{B-11e}), and the value \cite{BBNS-01}
\beq
\label{B-12e}
C_{7\gamma}^{eff}(m_b) = -0.318
\eeq

\noi we find for the Penguin contributions at $\mu = m_b$,  
$$P_4^{-,u} = -0.0086 - 0.0131 i \ , \qquad \qquad P_4^{-,c} = -0.0067 + 0.0012 i$$
\beq
\label{B-12e}
P_7^{-,u} = P_9^{-,u} = 0.0073 \ , \qquad \qquad  P_7^{-,c} = P_9^{-,c} = 0.0071
\eeq
$$P_{10}^{-,u} = P_{10}^{-,c} = (- 7. - 2. i)\times  10^{-5}$$

\vskip 30 truemm

\noi {\large \bf Hard scattering corrections} 

\vskip 5 truemm

The Hard scattering corrections and the Weak annihilation contributions are estimated at the scale $\mu = \sqrt{m_b\Lambda_h} = 1.45$. The Hard scattering contributions $H_i^-$ in (\ref{B-1e}) are given by the expressions
$$H_i^- = - {18 f_{B_s} f_\phi^\perp(\mu_h) \over m_{B_s} m_b F_-^{B_s \to \phi}(0) } {m_b \over \lambda_{B_s} }\ (H_H-1)\ , \qquad \   i \in \{1, 2, 3, 4, 9, 10 \}$$
\beq
\label{B-13e}
\ \ \ \ \ H_i^- = {18 f_{B_s} f_\phi^\perp(\mu_h) \over m_{B_s} m_b F_-^{B_s \to \phi}(0) } {m_b \over \lambda_{B_s} }\ (H_H-1)\ , \qquad \qquad \qquad \ \ \ \  i \in \{5, 7 \}
\eeq
$$H_i^- = - {9 f_{B_s} f_\phi \over m_{B_s} m_b F_-^{B_s \to \phi}(0) } {m_b \over m_\phi} {m_b \over \lambda_{B_s} }\ \ , \qquad \qquad \qquad  \qquad \ \  i \in \{6, 8 \}$$

\vskip 3 truemm

Let us make a few remarks on these formulas.\par
The quantities $H_i^-$ are a piece of the coefficients $a_i^-$ (\ref{B-1e}) inversely proportional to the form factor $F_-^{B_s \to \phi}(0)$. Notice that this gives a form-factor dependence of the coefficients $a_i^-$, the rest of the terms in (\ref{B-1e}) being form factor independent.\par  
The parameter $f_\phi^\perp$ is scale-dependent. At $1\ {\rm GeV}$ one has \cite{BUCHALLA} $f_\phi^\perp(1{\rm GeV}) = 0.186\ {\rm GeV}$, and we need this parameter at the scale $\mu_h = 1.45\ {\rm GeV}$,  
\beq
\label{B-13e}
f_\phi^\perp(\mu_h) = f_\phi^\perp(1{\rm GeV}) \left({\alpha_s(\mu_h) \over \alpha_s (1\ GeV)} \right)^{C_F/\beta_0} = 0.179\ {\rm GeV}
\eeq
\noi Finally, we take for the parameter $\lambda_{B_s} = 0.2\ {\rm GeV}$ and $H_H$ given by (\ref{C-9-1e}) \cite{BENEKE-2007}.\par
Numerically one finds,
$$H_i^- = - 1.522\ , \qquad \qquad \qquad \   i \in \{1, 2, 3, 4, 9, 10 \}$$
\beq
\label{B-14e}
H_i^- = 1.488\ , \qquad \qquad \qquad \qquad \qquad \ \ \ \ i \in \{5, 7 \}
\eeq
$$H_i^- = - 2.816\ , \qquad \qquad \qquad  \qquad \qquad\ \  i \in \{6, 8 \}$$

\vskip 3 truemm

\noi {\large \bf Summary on the coefficients $a_i^-$}

\vskip 3 truemm

Gathering the precedent results in (\ref{B-1e}), and taking care of the two different scales $\mu = m_b$ and $\mu_h = \sqrt{\mu \Lambda_h}$ with $\Lambda_h = 0.5\ {\rm GeV}$ for the different terms, one finds the coefficients $a_i^-$ of Table 7.

\vskip 20 truemm

\noi {\Large \bf The annihilation coefficients $b_i^-$}

\vskip 3 truemm

These coefficients are given by
$$b_3^- = {C_F \over N_c^2} \left[C_3 A_1^{i-} + C_5 (A_3^{i-} + A_3^{f-}) +N_c C_6 A_3^{f-} \right] \ \ \ \ $$
\beq
\label{B-15e}
b_4^- = {C_F \over N_c^2} \left[C_4 A_1^{i-} + C_6 A_2^{i-} \right] \qquad \qquad \qquad \qquad \ \ \ \ \ 
\eeq
$$b_3^{EW-} = {C_F \over N_c^2} \left[C_9 A_1^{i-} + C_7 (A_3^{i-} + A_3^{f-}) +N_c C_8 A_3^{f-} \right]$$
$$b_4^{EW-} = {C_F \over N_c^2} \left[C_{10} A_1^{i-} + C_8 A_2^{i-} \right] \qquad \qquad \qquad \qquad $$

\noi with \cite{BENEKE-2007}

$$A_1^{i-} = A_2^{i-} = 18 \pi \alpha_s {m_\phi^2 \over m_{B_s}^2} \left({1 \over 2} X_L + {5 \over 2} - {\pi^2 \over 3} \right)$$
\beq
\label{B-16e}
A_3^{i-} = 0 \qquad \qquad  \qquad \qquad \qquad \qquad \qquad \ \ 
\eeq
$$A_3^{f-} = 36 \pi \alpha_s r_\perp^\phi (2 X_A^2-5X_A+3) \qquad \qquad$$

\noi where the running of $r_\perp^\phi(\mu)$ is given by \cite{BUCHALLA} 
\beq
\label{B-17e}
r_\perp ^\phi(\mu) = {2m_\phi f_\phi^\perp(\mu) \over m_b(\mu) f_\phi} = {2m_\phi f_\phi^\perp(1 {\rm GeV}) \over m_b(m_b) f_\phi} \left[ {\alpha_s(\mu) \over \alpha_s(m_b)} \right]^{-3C_F/\beta_0} \left[ {\alpha_s(\mu) \over \alpha_s(1 {\rm GeV})} \right]^{C_F/\beta_0}
\eeq

\noi and must be computed at the scale $\mu_h$.

\vskip 3 truemm
\noi From $f_\phi^\perp(1 {\rm GeV}) = 0.186\ {\rm GeV}$ \cite{BUCHALLA} or $f_\phi^\perp(2 {\rm GeV}) = 0.175\ {\rm GeV}$ \cite{BENEKE-2007}, one finds consistently, for $\mu = \sqrt{m_b\Lambda_h} = 1.45\ {\rm GeV}$
\beq
\label{B-18e}
r_\perp ^\phi(1.45\ {\rm GeV}) = 0.318
\eeq 

\noi and $X_A$ and $X_L$ are given by (\ref{C-9-1e}).\par

Numerically one finds,
\beq
\label{B-19e}
A_1^{i-} = A_2^{i-} = 5.888 \ , \qquad A_3^{i-} = 0 \ , \qquad A_3^{f-} = 97.450 - 98.673 i
\eeq

\noi and therefore,
\beq
\label{B-20e}
b_3^- = -3.527 + 3.596 i \ , \qquad b_4^- = -0.129
\eeq
$$b_3^{EW-} = 0.034 - 0.044 i \ , \qquad b_4^{EW-} = 0.004$$

\noi that are given in Table 8, and one sees that the annihilation is largely dominated by $b_3^-$.

\vskip 10 truemm

\noi {\Large \bf Appendix C. The annihilation contribution to the amplitude $A(\overline{B}_s \to \phi \phi, h = 0)$}

\vskip 4 truemm

For the longitudinal amplitude, the coefficients $b_3^0, b_4^0, b_3^{EW0}, b_4^{EW0}$ are given by the expressions (\ref{B-15e}) with $h = - \to h = 0$ and \cite{BUCHALLA}
\beq
\label{C-1e}
A_1^{i0} = A_2^{i0} \ , \qquad \qquad A_3^{i0} = 0
\eeq

\noi and 
$$A_1^{i0} \simeq 18 \pi \alpha_s \left[\left(X_A - 4 + {\pi^2 \over 3} \right) + (r_\perp^V)^2 (X_A-2)^2\right]$$
\beq
\label{C-2e}
A_3^{f0} \simeq - 36 \pi \alpha_s r_\perp^V (2X_A^2-5X_A + 2)
\eeq

\noi With the parametrization
\beq
\label{C-3e}
X_A = \int_0^1 {dx \over x} = \left (1 + \rho_A e^{i \phi_A} \right) \ln {m_B \over \Lambda_h}
\eeq

\noi and the values $\rho_A = 0.6, \phi_A = - 40^0$ of ref. \cite{BENEKE-2007}, one finds the annihilation coefficients for the longitudinal amplitude of Table 3 for model (2). One finds a very large coefficient $b_3^0$, comparable in absolute magnitude to the transverse one $b_3^-$.

\vskip 10 truemm

\noi {\Large \bf Acknowledgements}

\vskip 5 truemm

We are very much indebted to Martin Beneke and Gerhard Buchalla for providing us useful and detailed information on the QCD Factorization scheme applied to the decays of $B$ mesons into two light vectors mesons.

\enddocument